\documentclass[twocolumn]{svjour3}

\RequirePackage[T1]{fontenc}

\smartqed  % flush right qed marks, e.g. at end of proof

\RequirePackage{graphicx}
\RequirePackage{mathptmx}      % use Times fonts if available on your TeX system
\RequirePackage{flushend}
\RequirePackage[numbers,sort&compress]{natbib}
\RequirePackage[colorlinks,citecolor=blue,urlcolor=blue,linkcolor=blue]{hyperref}

\RequirePackage[utf8]{inputenc}
\RequirePackage{amssymb}
\RequirePackage{amsmath}
\RequirePackage{braket}
\RequirePackage{xcolor}

\journalname{Eur. Phys. J. B}

\begin{document}

\title{Real and imaginary energy gaps: a comparison between single excitation Superradiance and Superconductivity and robustness to disorder}

\author{Nahum C. Ch\'avez
\and
Francesco Mattiotti
\and
J. A. M\'endez-Berm\'udez
\and
Fausto Borgonovi
\and
G. Luca Celardo
}

\institute{Nahum C. Ch\'avez \and J. A. M\'endez-Berm\'udez \and G. Luca Celardo \at Benem\'erita Universidad Aut\'onoma de Puebla, Apartado Postal J-48, Instituto de F\'isica,  72570, M\'exico
\and
Francesco Mattiotti \and Fausto Borgonovi \at Dipartimento di Matematica e Fisica and Interdisciplinary Laboratories for Advanced Materials Physics, Universit\`a Cattolica, via Musei 41, 25121 Brescia, Italy
\and
Francesco Mattiotti \and Fausto Borgonovi \at Istituto Nazionale di Fisica Nucleare,  Sezione di Pavia, via Bassi 6, I-27100,  Pavia, Italy
}

\date{\today}

\maketitle

\begin{abstract}
A comparison between the single particle spectrum of the discrete Bardeen-Cooper-Schrieffer (BCS) model, used for small superconducting grains, and the spectrum of a paradigmatic model of Single Excitation Superradiance (SES) is presented. They are both characterized by an equally spaced energy spectrum (Picket Fence) where all the levels are coupled between each other by a constant coupling which is real for the BCS model and purely imaginary for the SES model.  While the former corresponds to the discrete BCS-model describing the coupling of Cooper pairs in momentum space and it induces a Superconductive regime, the latter describes the coupling of single particle energy levels to a common decay channel and it induces a Superradiant transition. We show that the transition to a Superradiant regime can be connected to the emergence of an imaginary energy gap, similarly to the transition to a Superconductive regime where a real energy gap emerges. Despite their different physical origin, it is possible to show that both the Superradiant and the Superconducting gaps have the same magnitude in the large gap limit. Nevertheless, some differences appear: while  the critical coupling at which the Superradiant gap appears is  independent of the system size $N$, for the Superconductivity gap it scales as $(\ln N)^{-1}$. 
The presence of a gap in the imaginary energy axis between the Superradiant and the Subradiant states shares many similarities with the ``standard'' gap on the real energy axis: the superradiant state is protected against disorder from the imaginary gap as well as the superconducting ground state is protected by the real  energy gap. Moreover we connect the origin of the gapped phase to the long-range nature of the coupling between the energy levels. 
\end{abstract}

\section{Introduction}

Cooperative effects, which are at the basis of emergent properties~\cite{moreisdifferent}, are 
at the center of research investigations in a
vast variety of fields: emergent properties in highly
correlated materials~\cite{dagotto}, cooperative emission in
superconducting qubits~\cite{qed}, Superradiance in cold atomic clouds~\cite{kaiser}, cooperative shielding in long range interacting systems~\cite{lea},
collective excitations in semiconductors~\cite{chrea}, plasmonic
Dicke effect~\cite{plasmons}, biophysical systems~\cite{schulten,srphoto2} and proposal of quantum devices which exploits cooperative effects~\cite{superab}. 
Despite the great importance of emergent properties,
a general unifying framework and a full understanding of cooperative effects
has not been found yet.
One of the most interesting properties of cooperative
effects is their robustness to the noise induced by external
environments. A well known example is Superconductivity, but 
other quantum emergent effects, such as 
Single Excitation Superradiance (SES),
have also been shown to be robust to
noise~\cite{Giulio,alberto}. This suggests that emergent properties could play an essential role in
the successful development of scalable quantum devices able to operate
at room temperature. 
%If built, they will outperform classical devices currently used for applications such as computing and light-harvesting.
Since cooperative effects represent a common mechanism to all these emergent phenomena,
we believe that finding links between different cooperative effects will be 
fundamental to progress our understanding of emergence.
%a fundamental progress in their understanding  Finding links between different cooperative effects, a common mechanism to emergence, further evidence of cooperativity in realistic systems, will be fundamental to progress in our understanding of emergence.
As was suggested by U. Fano~\cite{fano} a common mechanism underlies
several collective phenomena, such as Superconductivity, plasmon excitation and giant resonances in nuclei.
In particular a possible connection between Superradiance and
Superconductivity has been discussed by M. Scully~\cite{scully}.

\begin{figure}[ht!]
  \centering
  \includegraphics[width=0.95\columnwidth]{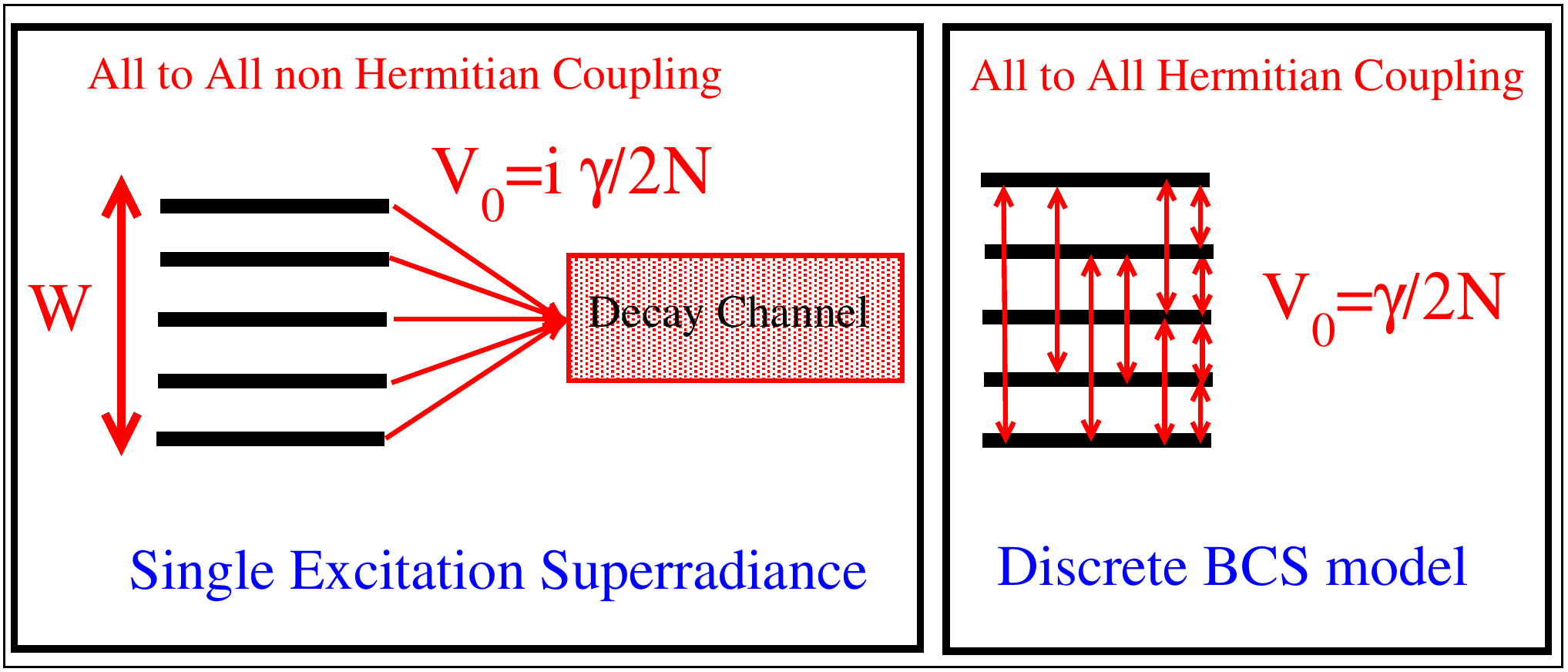}
    \includegraphics[width=0.95\columnwidth]{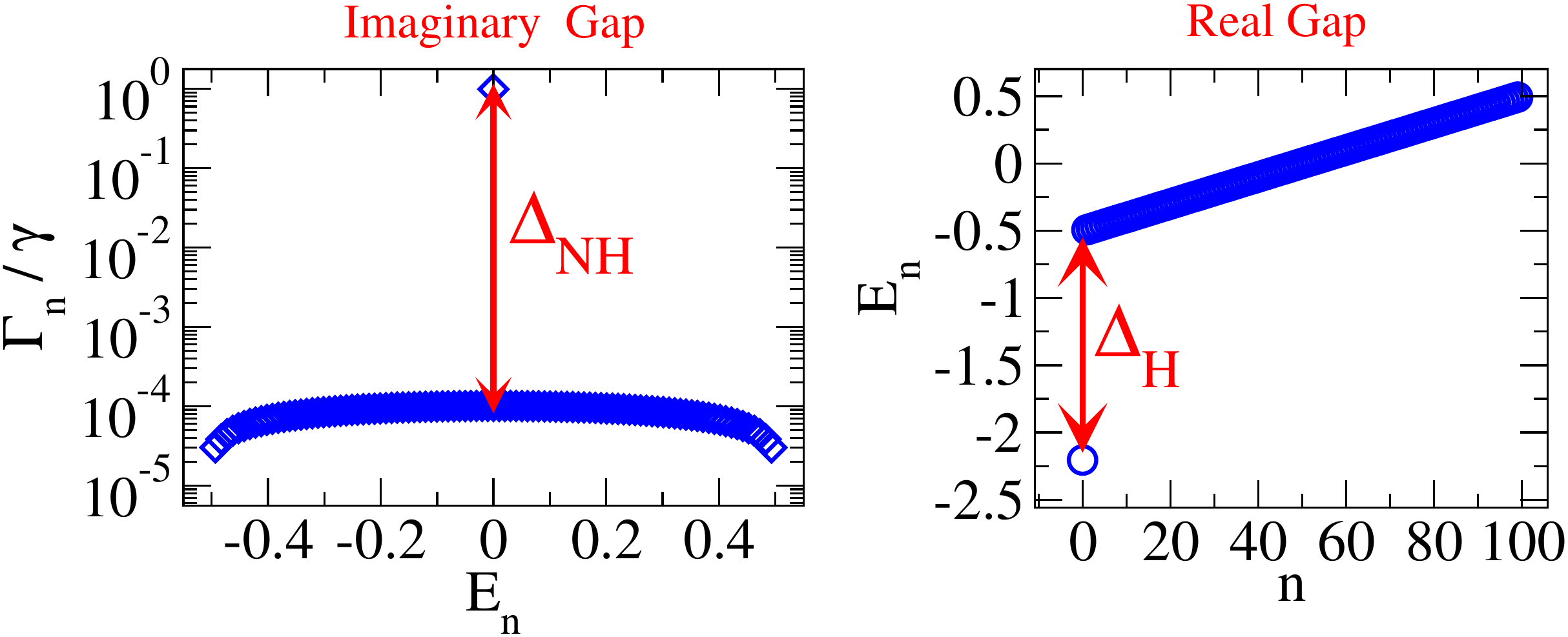}
    \caption{(Left upper panel) Paradigmatic model of SES: the coupling of equally spaced single particle energy levels to a common channel in the continuum induces a non-Hermitian all to all interaction between the energy levels described by the Hamiltonian in Eq.~\eqref{H} with $V_0=i\gamma/(2N)$. (Right upper panel) The discrete BCS model of Superconductivity where Cooper pair states are coupled by an all to all hermitian interaction, see arrows between levels, described by the Hamiltonian in Eq.~\eqref{H} with $V_0= \gamma/(2N)$. (Lower left panel) Complex eigenvalues $E_n-i\Gamma_n/2$ for the SES model and the imaginary energy gap are shown. (Lower right panel) The eigenvalues $E_n$ and the real energy gap are shown for the BCS model. Parameters are: $N=100$, $W=1$, $\gamma=10 \gamma_{\rm cr}^{\rm H,NH}$, where $\gamma_{\rm cr}^{\rm H,NH}$ is the critical coupling for the BCS/SES transition.}
  \label{f1}
\end{figure}

Here we perform a comparison between Superconductivity, i.e. the discrete Bardeen-Cooper-Schrieffer (BCS) model, and Single Excitation Superradiance (SES model). We show that in both cases we have the emergence of a ``gap" in the energy spectrum.
Superradiance is usually referred to the case of many excitations in an ensemble of $N$ two level systems and to the existence of states which emit energy with an intensity proportional to $N^2$. On the other hand, SES refers to the possibility that a single excitation coherently shared by $N$ two level systems can decay with a rate proportional to $N$, an effect defined as the Super of Superradiance in Ref.~\cite{super} due to the fact that SES involves a fully entangled state.
%Single excitation Super and Sub-radiance has been found experimentally
%in cold atomic clouds, and it is thought to have an important role in photosynthetic systems. 

Specifically,  we analyze a paradigmatic model of SES, which has been studied in~\cite{rotter}, see Fig.~\ref{f1}.  
In such model, the single excitation energy levels are assumed equally spaced and connected to a single decay channel in the continuum. Due to the fact that the system is open and the excitation can be lost in the common decay channel, the eigenvalues of the system are complex. When the resonances overlap, a Superradiance transition occurs: a Superradiant state  acquires most of the decay width of the system, while the other $N-1$ subradiant states decrease their own widths on increasing the coupling strength with the common decay channel. In the limit of large coupling to the continuum only the Superradiant state can decay. 
Here we show that the Superradiance transition is connected with the emergence of an imaginary energy gap between the complex eigenvalues of the system. 
% The energy gap is complex since  the 
%states can decay and thus  they are characterized by a complex
%energy: ${\cal E}= E-i\Gamma/2$, where $\Gamma$ is their decay width.  
Our aim is to investigate and compare the energy gaps arising
in such paradigmatic
model of Superradiance~\cite{rotter,Zannals} with the well-known energy gap
present in a 
model of Superconductivity (the discrete BCS model~\cite{bcs,bcs2}), paying main attention to the robustness to disorder induced by the presence of a gap and to the kind of interaction which originates the gap.

The discrete BCS model is widely used to analyze Superconductivity in
small metallic grains~\cite{bcs}. Moreover, the single particle (single Cooper pair) sector of the discrete BCS model had been studied in several papers~\cite{bcs2}. In this case, the model  is very similar to the model proposed by
L. Cooper in his seminal paper~\cite{cooper} and its Hamiltonian reads: 
\begin{equation}
  H = H_0 + V = \sum_k E^0_k \ket{k} \bra{k} - V_0 \sum_{k,k'} \ket{k} \bra{k'} \, ,
\label{H}
\end{equation}
where $\ket{k}$ is the Cooper pair state, $E^0_k$ is the unperturbed
energy, usually taken as equally spaced (picket fence (PF) spectrum), and $V_0$ is the coupling
between the Cooper pair states. The coupling is the same for
all the states, similarly to what happens in models with an infinite range
coupling in space, with the difference that here the coupling is in
momentum space.  
The same model, see Fig.~\ref{f1},
is used to describe the Superradiance transition in a
system where many levels are coupled to the same channel in the
external environment~\cite{rotter}. The only but important difference
is that for the case of Superradiance $V_0$ is a pure imaginary
number. In this case $\ket{k}$ represents a single energy
level or an atomic or molecular excitonic state in a specific point
of the real space. For the case of Superradiance, the non-Hermitian
Hamiltonian, originating from the imaginary coupling,
takes into account the fact that the system
can decay into the continuum but it also represents the
coupling between the energy levels which modifies the spectral
features of the system~\cite{sr}. The limit of validity and the effectiveness of the effective non-Hermitian Hamiltonian description of the system has been investigated in Ref.~\cite{NHH}. In such systems a transition to Superradiance
occurs above a critical coupling strength. In the Superradiant
regime, when one  Superradiant state acquire most of the decay
width of the system, a gap opens in the complex energy plane of  the non-Hermitian Hamiltonian, see Fig.~\ref{f1} lower left panel. 

Note that the model presented in Eq.~\eqref{H}, apart from being relevant in studying cooperative effects, it is also relevant in describing realistic systems. For instance, for the hermitian case, this model can be reproduced in ion trap experiments with a tunable interaction range, including all-to-all coupling~\cite{iontraps,robincs}. On the other side, for the non-Hermitian case, this model is relevant in nuclear physics~\cite{sr} and it could be also deviced in molecular systems~\cite{srphoto2}.

%Here we compare the imaginary gap present in the SES model with the real gap emerging in the discrete BCS model. We show that the Superradiance transition coincides with the opening of a gap in the complex energy plane. In the limit of large gap, the magnitude of the Superconducting gap is the same as the Superradiance gap. Nevertheless, our analytical results show that the critical coupling which determines the Superradiance transition scales differently, with the system size, from that for the Superconductivity transition.
 
%Finally, using a perturbative approach (up to second order) for the non-Hermitian Hamiltonian, we show that such complex energy gap can protect the states from perturbations in the same way as a gap in real energy spectra. This result is consistent with several results found in literature~\cite{alberto,robincs,Giulio} about the robustness of Superradiance to disorder. 

In Sec.~\ref{secAnalytical} we compare the imaginary gap present in the SES model with the real gap emerging in 
the discrete BCS model. We show that the Superradiance transition coincides with the opening of a gap in the complex energy plane. 
In the limit of large gap, the magnitude of the Superconducting gap is the same as the Superradiance gap. Nevertheless, 
our analytical results show that the critical coupling which determines the Superradiance transition scales with the system size differently 
 from the critical coupling associated to the Superconductivity transition. In Sec.~\ref{secNumerical} we present a few numerical results 
showing the validity of our analytical equations. Then, in Sec.~\ref{secRob} we apply the perturbation theory (up to second order) to 
non-Hermitian systems and we show, also through numerical simulations, how the imaginary energy gap can protect the states from perturbations, such as  static disorder, 
in the same way as a real energy gap does. This result is consistent with several results found in literature~\cite{alberto,robincs,Giulio} 
about the robustness of Superradiance to disorder.
Finally, in Sec.~\ref{secAlpha} we analyze the role of the range of the interaction, showing that an energy gap emerges in the Hermitian system when the interaction is long-ranged.   
In the Conclusions the relevance of our analysis to realistic systems is  discussed. 

\section{Analytical Results for $N$ levels}
\label{secAnalytical}

Let us consider $N$  equally spaced levels in an energy range $W$,
coupled between each other with a constant coupling $V_0$, which can be real or
imaginary. The Hamiltonian can thus be written as in Eq.~\eqref{H},
where for the energy we assume a PF distribution, namely
\begin{equation}
  \label{pf}
  E^0_k = k\delta = k\frac{W}{N} \, , \qquad k = -\frac{N}{2}, \dots, \frac{N}{2} \, ,
\end{equation}
where $\delta=W/N$ is the level spacing. First, for the sake of clarity, we present the derivation of the Gap
Equation~\cite{cooper,schrieffer,rotter}, both for the Hermitian and
non-Hermitian cases, which is equivalent to the
Schr\"{o}dinger Equation and it makes the computation of some eigenvalues
much easier. For the derivation of the Gap Equation we follow~\cite{schrieffer}, which presents a
simplified version of the famous derivation by L. Cooper in his
seminal paper~\cite{cooper}. 
%Our case is more simple because we neglect the kinetic energy of the nuclei.

We want to solve the Schr\"{o}dinger Equation
\begin{equation}
  \label{schr}
  H \ket{\Psi} =E \ket{\Psi} ,
\end{equation}
where $\ket{\Psi}$ is an eigenstate of the full Hamiltonian $H$ and it can be expanded as
\begin{equation}
  \label{psi}
  \ket{\Psi} = \sum_k a_k \ket{k} , 
\end{equation}
where $\ket{k}$ are eigenstates of $H_0$, satisfying
\begin{equation}
  H_0 \ket{k} =E^0_k \ket{k} \, .
\end{equation}
Eq.~\eqref{schr} can be rewritten as
\begin{equation}
  (H-H_0)\ket{\Psi}=V \ket{\Psi} \, .
\end{equation}
Then, using the expansion~\eqref{psi}, we get
\begin{equation}
\label{eq-ppp}
  (H-H_0)\sum_{k'} a_{k'} \ket{k'} =V \sum_{k'} a_{k'} \ket{k'} =\sum_{k'} a_{k'} V\ket{k'} \, .
\end{equation}
Now let us project Eq.~\eqref{eq-ppp} on the state $\bra{k}$. Defining $V_{kk'}=\braket{k|V|k'}$ we have
\begin{equation}
\bra{k} \sum_{k'} (E-E^0_{k'}) a_{k'} \ket{k'} =\sum_{k'} a_{k'} V_{kk'} \, .
\end{equation}
Now, since $\braket{k|k'} = \delta_{k,k'}$ and $V_{kk'}=-V_0 \ \forall k,k'$  we get
\begin{equation}
  \sum_{k'} (E-E^0_{k'}) a_{k'} \braket{k|k'} =(E-E^0_k) a_k = -V_0 \sum_k a_k \, .
\end{equation}
Defining $C \equiv \sum_k a_k$ we have
\begin{equation}
a_k=- \frac{V_0 C}{E-E^0_k}
\end{equation}
so that
\begin{equation}
C = \sum_k a_k=-V_0 \sum_k \frac{C}{E-E^0_k} \, .
\end{equation}
Dividing by $C$ we finally obtain the Gap Equation
\begin{equation}\label{GapEq}
  1 = -V_0 \sum_k \frac{1}{E-E^0_k} \, .
\end{equation}
Eq.~\eqref{GapEq} has been obtained by simple linear manipulations of the Schr\"{o}dinger equation, and so they are equivalent. Given the unperturbed eigenvalues $E^0_k$, there are $N$ possible values of $E$ which satisfy Eq.~\eqref{GapEq}, which are the eigenvalues of $H$.

The term Gap Equation comes from the fact that it is commonly used to compute the gap between the ground state and the excited states. In the next sections we will compute the gap
 for the case  $V_0=\gamma/(2N)$ real (for Superconductivity) and for  $V_0=i\gamma/(2N)$ complex (for Superradiance). Note that in both cases we rescale the coupling by $N$ as it is found in the discrete BCS model~\cite{bcs,bcs2}). The non rescaled case can be easily deduced by substituting $\gamma$ with $N\gamma$ in the following results.

\subsection{Hermitian case}

\begin{figure}[t]
  \centering
  \includegraphics[width=\columnwidth]{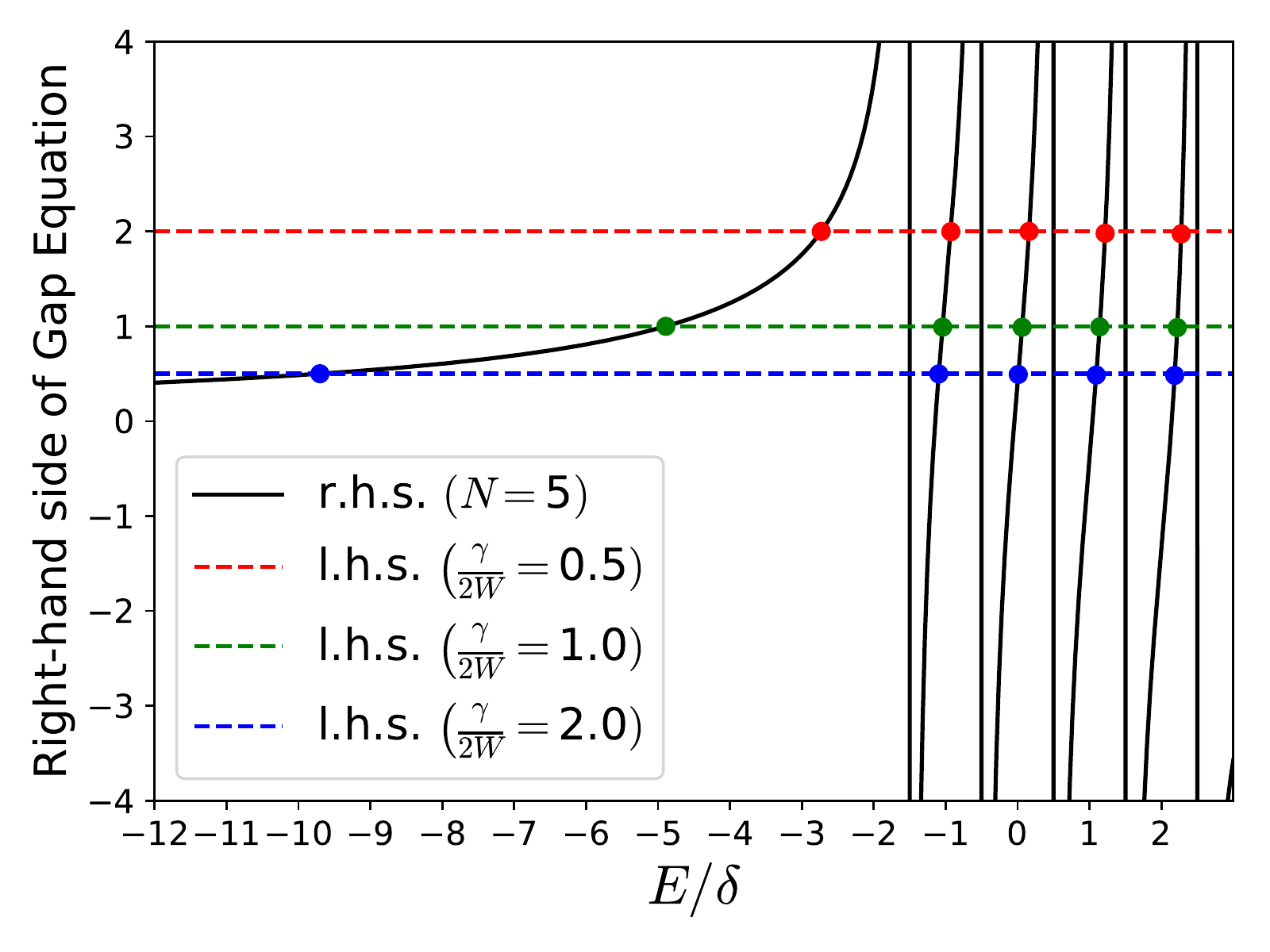}
  \caption{Graphical solution of Gap Equation~\eqref{gapeqh} in the Hermitian case. The all-to-all coupling here is $V_0=\gamma/(2N)$. We set $N=5$.}
  \label{gapeqf}
\end{figure}

Let us review the main results about the gap equation for the Hermitian case, see also Refs.~\cite{bcs2,cooper,schrieffer}. 
Let us now consider the Hermitian case, with $V_0=\gamma/(2N)$. Recalling that the unperturbed spectrum is given by~\eqref{pf}, we multiply both sides of~\eqref{GapEq} by $2W/\gamma$ to have
\begin{equation}
  \label{gapeqh}
  \frac{2W}{\gamma} = \sum_{k=-N/2}^{N/2} \frac{1}{k - E/\delta} \, .
\end{equation}
A graphical solution of Eq.~\eqref{gapeqh} is shown in Fig.~\ref{gapeqf} for $N=5$. The r.h.s. of Eq.~\eqref{gapeqh}, shown as a continuous black line, is an unbounded function of $E/\delta$ having $N$ asymptotes, corresponding to $E/\delta=-N/2, \dots, N/2$ and shown as vertical black lines. The l.h.s., shown as dashed lines, is independent of $E/\delta$. The solutions are given by the values of $E$ where the r.h.s. intersects the l.h.s. and they are shown by full circles in Fig.~\ref{gapeqf}.

From Fig.~\ref{gapeqf} one can see that there are $N=5$ solutions to Eq.~\eqref{gapeqh}. Those solutions represent the eigenvalues $E$ of the full Hamiltonian~\eqref{H}, divided by the level spacing $\delta=W/N$ of the unperturbed levels. Moreover, one can observe that, by increasing the ratio $\gamma/(2W)$ (so to  increase the all-to-all coupling), the energy gap between the ground state and the first excited state increases, too. We are here interested in computing that gap in the limit $N \to \infty$ and keeping $W = \text{const.}$, so that the spacing $\delta$ of the unperturbed levels tends to 0. First of all, from Fig.~\ref{gapeqf} we can see that the energy of the excited states are all in the range $[-W/2,W/2]$ and, in particular, the first excited state lies in the interval $-W/2+\delta < E_2 < -W/2+2\delta$. This implies that only the ground state energy $E_1$  can be less than $-W/2$ and that the energy of the first excited state $E_2$ tends to $-W/2$ when $\delta \to 0$. Now, let us focus on the energy of the ground state.

If $N \gg 1$, we can take the continuum limit for the Gap Equation~\eqref{GapEq}
\begin{equation}
  1=-\frac{\gamma}{2N} \int_{-W/2}^{W/2} \frac{N(x) dx}{E-x} ,
\end{equation}
where $N(x)=N/W$ is the density of states and it is constant for a PF level distribution. Then we can analytically solve the integral
\begin{equation}
  1= -\frac{\gamma}{2W} \int_{-W/2}^{W/2} \frac{dx}{E-x}=\frac{\gamma}{2W} \ln \frac{2E-W}{2E + W} \, ,
\end{equation}
noting that the above solution is valid only for $E<-W/2$. For what we stated before, the only state which satisfies this requirement is the ground state $E_1$. Then we have
\begin{equation}
E_1=\frac{W}{2}\frac{1+e^{2W/\gamma}}{1-e^{2W/\gamma}} \, .
\label{GS}
\end{equation}

Now, let us recall that  $E_2 \to -W/2$ when $N \to \infty$ and $W$
does not depend on $N$ (see~\cite{schrieffer} and the previous considerations on Fig.~\ref{gapeqf}). We can then define the Hermitian Gap between the ground state and the first excited state as
\begin{equation}
  \Delta_{\rm H}=E_2 - E_1 = -\frac{W}{2}-E_1=\frac{W}{e^{2W/\gamma}-1} \, .
\label{gaph}
\end{equation}

Note that the expression of the gap  obtained is the same as
the one obtained by L. Cooper~\cite{cooper}, but it is slightly different
from the BCS gap~\cite{bcs,bcs2,note},
In the limit $W \ll \gamma/2$ Eq.~\eqref{gaph}  is approximated as
\begin{equation}
\label{gap-hl}
 \Delta_{\rm H} \approx \frac{\gamma}{2} \, .
\end{equation}
On the other hand, when $W \gg \gamma/2$, Eq.~\eqref{gaph} becomes
\begin{equation}
    \Delta_{\rm H} \approx W \, e^{-2W/\gamma} \\
\end{equation}

On increasing $\gamma$, the gap $\Delta_{\rm H}$ increases as well and, for some
$\gamma=\gamma_{\rm cr}^{\rm H} $ it becomes equal to the unperturbed level spacing
$\delta=W/N$. By setting $\Delta_{\rm H}=\delta$, it is easy to find that
for $N \gg 1$,
\begin{equation}
  \label{gcrh}
  \gamma_{\rm cr}^{\rm H}  = \frac{2W}{\ln(N+1)} \approx \frac{2W}{\ln N},
\end{equation}
 which defines the critical coupling at which a gap opens in the BCS model.
 
\subsection{Non-Hermitian case}

\begin{figure}[ht]
  \centering
  \includegraphics[width=\columnwidth]{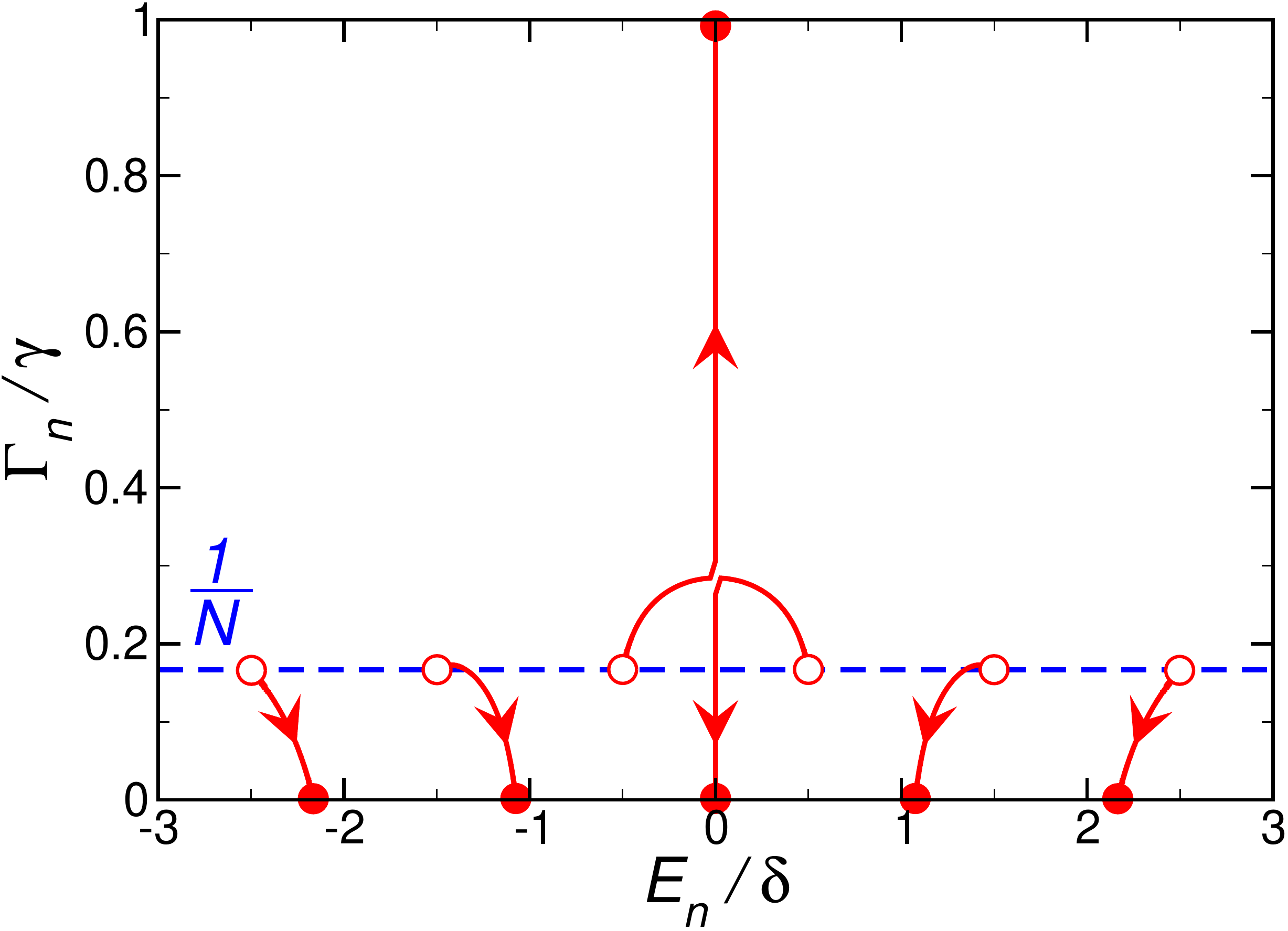}
  \includegraphics[width=\columnwidth]{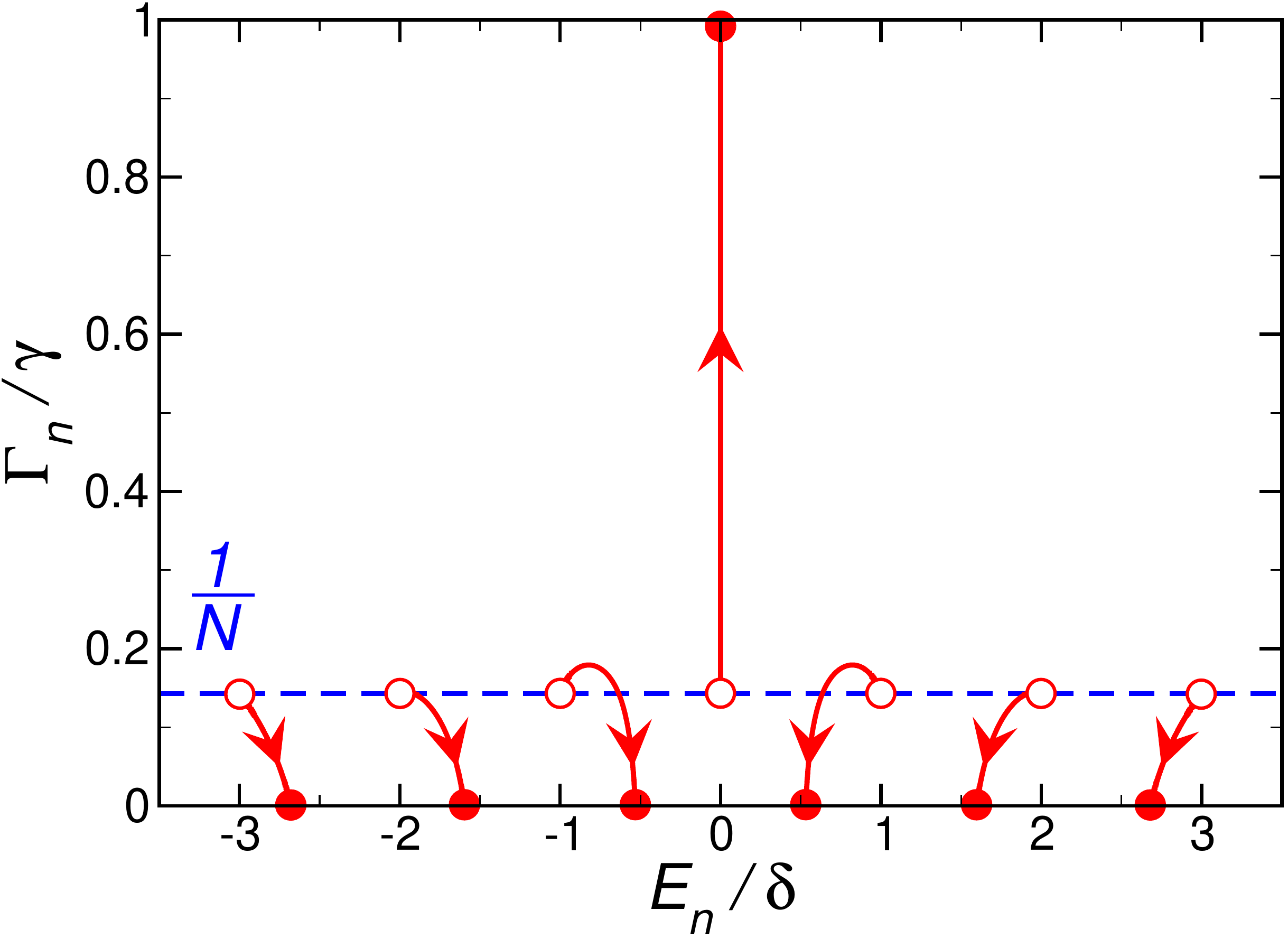}
  \caption{Complex eigenvalues~\eqref{ecomp} varying $\gamma$ from $\gamma=(2W/\pi)/10$ (open circles) to $\gamma=10(2W/\pi)$ (full circles). $\gamma$ increases following the arrows. The dashed line marks the value $\Gamma_n=\gamma/N$. Parameters: $N=6$ (upper panel) and $N=7$ (lower panel).}
  \label{widthsf}
\end{figure}

\paragraph{Superradiant state (Gap Equation)} Now, let us consider the non-Hermitian case $V_0=i\gamma/(2N)$. Starting from Eq.~\eqref{GapEq} we obtain a gap equation
\begin{equation}
  1 = -\frac{i\gamma}{2N} \sum_k \frac{1}{\mathcal{E}-E^0_k} \, ,
\end{equation}
where the eigenvalues are now complex,
\begin{equation}
  \label{ecomp}
  \mathcal{E} = E - i\frac{\Gamma}{2} \, .
\end{equation}
This complex equation splits into two real equations
\begin{equation}
  \begin{cases}
    \displaystyle \sum_k \frac{E-E^0_k}{(E-E^0_k)^2+\Gamma^2/4} = 0 \\
    \displaystyle \sum_k \frac{\Gamma/2}{(E-E^0_k)^2+\Gamma^2/4} = \frac{2N}{\gamma} \label{gapeqnhim}
  \end{cases}
\end{equation}
which have $N$ solutions that depend on $N$ and $\gamma$. In
Fig.~\ref{widthsf} we plot the eigenvalues~\eqref{ecomp} in the plane
$(E/\delta,\Gamma/\gamma)$ for $N=6$ (upper panel) and $N=7$ (lower
panel), as a function of $\gamma$. In particular, we plot the trajectories of the eigenvalues starting from $\gamma=(2W/\pi)/10$ (open circles) up to $\gamma=10(2W/\pi)$ (full circles). The value $\gamma=2W/\pi$ marks the Superradiance transition, as we will show here below. When $\gamma$ is small (open circles in Fig.~\ref{widthsf}), the real parts of the eigenvalues are given by~\eqref{pf}, while the imaginary part is $\Gamma_n\approx\gamma/N$ for all eigenvalues. On increasing $\gamma$, the spacing between the real parts of the eigenvalues decreases (a phenomenon called ``pole attraction'') up to a critical point $\gamma_{\rm SR}$, where we see a different behaviour between $N=6$ and $N=7$.

For $N=6$ (upper panel of Fig.~\ref{widthsf}), the two eigenvalues whose real part is closer to 0 
collapse to the imaginary energy axis (so that $E_n=0$ for both of them) when  $\gamma > \gamma_{\rm SR}$. The widths of those two eigenvalues, however, have a different behaviour because one increases with $\gamma$ (and we call the respective  state \emph{Superradiant}) while the other one decreases with $\gamma$. For $\gamma \gg \gamma_{\rm SR}$ the total decay width of the system is concentrated in the superradiant state, while the other $N-1$ states have a negligible decay width, and thus they are called \emph{subradiant}.

For $N=7$ (lower panel of Fig.~\ref{widthsf}) the behaviour is similar  to that of $N=6$ in that, on increasing $\gamma$, the real parts of the eigenvalues are attracted to each other and for $\gamma \gg \gamma_{\rm SR}$ the total decay width of the system is concentrated into one superradiant state. The difference here (with respect to $N=6$) is that for $\gamma > \gamma_{\rm SR}$ only the Superradiant state has $E_{\rm SR}=0$, while $E_n \ne 0$ for the subradiant states.

The same behaviour seen for $N=6$ has been observed for all even values of $N$, while the behaviour observed for $N=7$ has been seen for all odd values of $N$. In the following calculations we look for an analytical expression for the width $\Gamma_{\rm SR}$ of the superradiant state and for the critical coupling $\gamma_{\rm SR}$ and, based on  the above discussion, we can set $E=0$ in Eq.~\eqref{gapeqnhim}.

Moreover, in the limit $N \gg 1$ we approximate the PF spectrum~\eqref{pf} with a continuous energy distribution constant in the interval $[-W/2,W/2]$, so that we can  solve the second equation in~\eqref{gapeqnhim}
\begin{equation}
  \frac{2N}{\gamma} = \frac{N}{W} \int_{-W/2}^{W/2} dx
  \frac{\Gamma/2}{x^2+\Gamma^2/4} = \frac{2N}{W} \arctan
  \frac{W}{\Gamma} \, ,
\label{e25}
\end{equation}
from which we get the width of the superradiant state
\begin{equation}
  \label{widthSR}
  \Gamma_{\rm SR} = \frac{W}{\tan \frac{W}{\gamma}} \, .
\end{equation}
This term is crucial to determine the gap in the
complex plane between the superradiant  and the closest subradiant
state. Note that $\Gamma_{\rm SR}$ has to be positive, and this gives
the condition of validity of Eq.~\eqref{e25}, which is $ \gamma \geq  \frac{2W}{\pi} \, .$
Therefore the superradiant state exists only above a critical coupling strength which coincides with the so-called Superradiance transition (at $\gamma=\gamma_{\rm SR}$), as we will show below.

\paragraph{Superradiant transition.}

In Ref.~\cite{rotter} the critical coupling at which a Superradiance transition occurs has been computed analytically by studying the dependence of the widths of the subradiant states on $\gamma$. Indeed below the Superradiance transition the widths of the subradiant states increase with $\gamma$, while above it, they decrease with $\gamma$. 
From Ref.~\cite{rotter} we have:
\begin{equation}
 \gamma_{\rm SR} = \frac{2W}{\pi} \, .
\label{gsr}
\end{equation}
Which is the same critical value of $\gamma$ computed in the previous section.

Adapting the analytical results of Ref.~\cite{rotter} to our case (see Appendix~\ref{appsub} for details) the decay widths of all the eigenstates below the Superradiance transition are
\begin{equation}\label{width3}
  \Gamma=\frac{W}{N\pi}\ln\left(\frac{1+\gamma/\gamma_{\rm SR}}{1-\gamma/\gamma_{\rm SR}}\right) \qquad \text{for } \gamma < \gamma_{\rm SR} \, ,
\end{equation}
while all the widths of the subradiant states above the Superradiance transition are
\begin{equation}\label{width4}
  \Gamma_{\rm sub}=\frac{W}{N\pi}\ln\left(\frac{\gamma/\gamma_{\rm SR} +1}{\gamma/\gamma_{\rm SR} -1}\right) \qquad \text{for } \gamma > \gamma_{\rm SR} \, .
\end{equation}
Note that the critical coupling parameter $\gamma_\mathrm{SR}$ is the point where the widths~\eqref{width3}-\eqref{width4} are non-analytical. 
% The non-analiticity at $\gamma=\gamma_{\rm SR}$ can be estimated (see~\cite{rotter} for details) as $\Gamma \sim (\ln N)/N$, so that the widths of the subradiant states tend to 0 for $N \to \infty$ for any value of $\gamma$, even for $\gamma=\gamma_{\rm SR}$.

\paragraph{Imaginary energy gap.}

The gap in the complex energy plane can be defined as 
\begin{equation}
  \Delta_{\rm NH} = \max_i \left\{ \min_{j \ne i}\left[ \text{dist} \left(\mathcal{E}_i, \mathcal{E}_j\right) \right] \right\} \, ,
  \label{d1}
\end{equation}
where the distance in the complex plane between two eigenvalues is
\begin{equation}
  \text{dist} \left( \mathcal{E}_i, \mathcal{E}_j\right) = \sqrt{\left( E_i-E_j\right)^2 + \frac{1}{4}\left( \Gamma_i-\Gamma_j\right)^2} \, .
  \label{d2}
\end{equation}
We can use the previous analytical results given in Eqs.~(\ref{width3},\ref{width4}) to estimate such complex gap. For  $\gamma < \gamma_{\rm SR}$, the widths of all the states are the same and the distance in real energy is constant and equal to $\delta$, where $\delta$ is the level spacing in the PF model, see Eq.~(\ref{pf}), so that we have $\Delta_{\rm NH}=\delta$ and no gap is present. On the other side 
in the superradiant regime $\gamma > \gamma_{\rm SR}$, we can estimate $\Delta_{\rm NH}$ as the distance in the complex plane between the superradiant eigenstate $\mathcal{E}_{\rm SR}$ and the closest subradiant state $\mathcal{E}_{\rm sub}$, see Appendix~\ref{appsub} for details, namely
\begin{equation}
  \Delta_{\rm NH} = \sqrt{\left( E_{\rm SR}-E_{\rm sub}\right)^2 + \frac{1}{4}\left( \Gamma_{\rm SR}-\Gamma_{\rm sub}\right)^2} \, .
\end{equation}
When $N \to \infty$ we have $\left( E_{\rm SR}-E_{\rm sub}\right) \approx \delta \to 0$ and $\Gamma_{\rm sub} \to 0$ (see~\eqref{width4}), so that the gap $\Delta_{\rm NH}$ is determined only by the decay width of the superradiant state~\eqref{widthSR}, 
\begin{equation}
  \label{gapnh}
  \lim_{N \to \infty} \Delta_{\rm NH} = \frac{\Gamma_\mathrm{SR}}{2} = \frac{W}{2\tan \frac{W}{\gamma}} \, .
\end{equation}
Now, we can define the critical value $\gamma_{\rm cr}^{\rm NH}$ as the value of $\gamma$ at which the gap opens, i.e. by imposing $\Delta_{\rm NH} = \delta$. From Eq.~\eqref{gapnh},  %a gap opens  when  $\Delta_{\rm NH} >0$, which occurs at the   critical value $\gamma_{\rm cr}^{NH}$: 
%\begin{equation}
%  \gamma_{\rm cr}^{NH} = %\gamma_{\rm SR} = \frac{2W}{\pi}.
%\end{equation}
we then have
\begin{equation}
    \label{gcrnh}
    \gamma_{\rm cr}^{\rm NH} = \frac{W}{\arctan \frac{N}{2}} \, .
\end{equation}
Note that this value is finite in the $N \to \infty$ limit,
\begin{equation}
  \label{gcrnhinf}
  \lim_{N \to \infty} \gamma_{\rm cr}^{\rm NH} = \frac{2W}{\pi} \, ,
\end{equation}
and it has the same value as $\gamma_{\rm SR}$ (see Eq.~\eqref{gsr}). Our results demonstrate that the Superradiant transition, previously analyzed in Ref.~\cite{rotter}, is equivalent to the emergence of a gap in the imaginary energy axis.

Moreover, we can approximate the gap for large $\gamma$, close to the transition and below the critical point, respectively, as
\begin{subequations}
  \begin{align}
    \Delta_{\rm NH} &\approx \frac{\gamma}{2} &\text{for } \gamma \gg \gamma_{\rm SR} \\
    \Delta_{\rm NH} &\approx \frac{\pi W}{4} \left( \frac{\gamma}{\gamma_{\rm cr}^{\rm NH}} - 1 \right) &\text{for } \gamma \gtrsim \gamma_{\rm SR} \\
    \Delta_{\rm NH} &= 0 &\text{for } \gamma \leq \gamma_{\rm SR} \, .
  \end{align}
\end{subequations}
Note that for $ \gamma \gg \gamma_{\rm SR} $ the complex energy gap of
the superradiant model is identical to the real energy gap of the
superconductivity model, see Eq.~\eqref{gap-hl}. On the other side, in the limit of large system sizes, the critical coupling for the emergence of a gapped state goes to zero for the BCS model, while for the SES model it remains finite.

\section{Numerical Results}
\label{secNumerical}

Here we validate our previous analytical predictions with few numerical results. 

\begin{figure*}[ht]
  \centering
  \includegraphics[width=\textwidth]{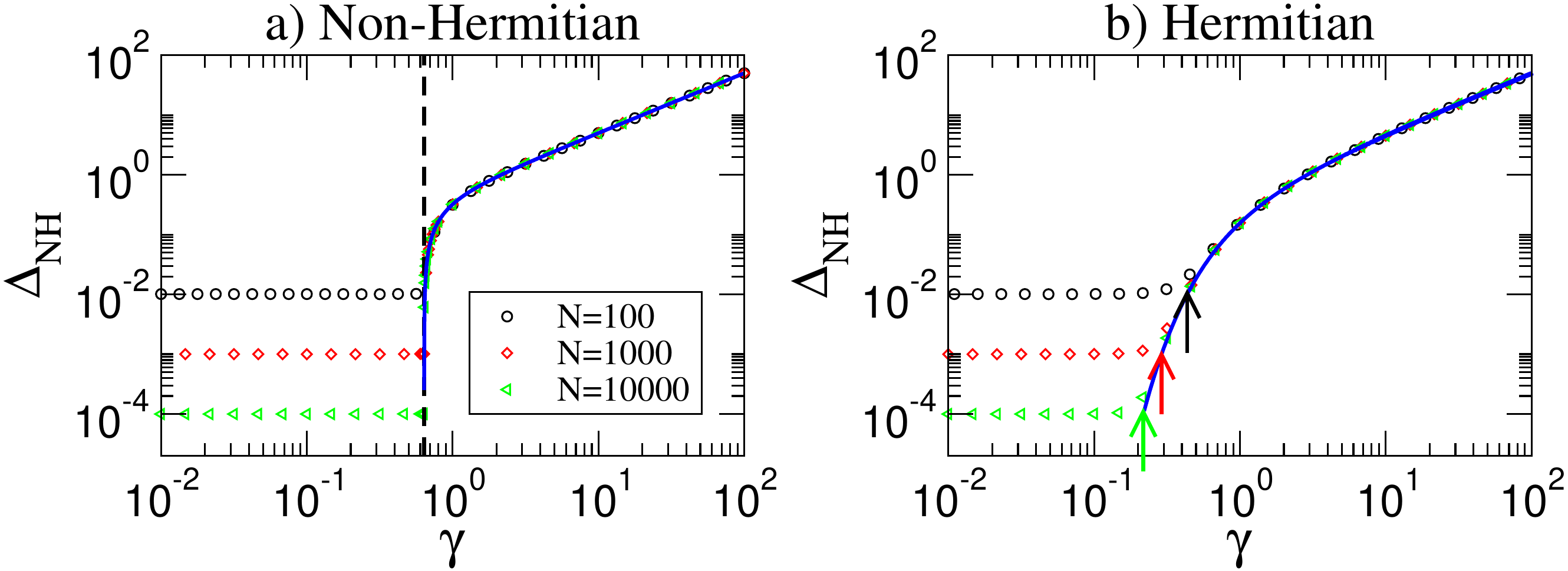}
  \caption{a) Gap $\Delta_{\rm NH}$ {\it vs.}~the coupling parameter $\gamma$. Symbols are given by~\eqref{d1}-\eqref{d2}, while the continuous blue curve shows the analytical estimate~\eqref{gapnh}. The dashed vertical line marks the critical coupling $\gamma_{\rm cr}^{\rm NH}$ from Eq.~\eqref{gcrnhinf}. b) Gap $\Delta_{\rm H}$ {\it vs.}~the coupling parameter $\gamma$. Symbols are given by~\eqref{gaph2} while the continuous blue curve shows the analytical estimate~\eqref{gaph}. The arrows indicate the critical coupling $\gamma_{\rm cr}^{\rm H}$ from Eq.~\eqref{gcrh}. Here, $W=1$ and $N=\{100,1000,10000\}$.}
  \label{Fig:Gap}
\end{figure*}

In Fig.~\ref{Fig:Gap} the gap is shown, both
for the Hermitian and non-Hermitian cases, as a function of $\gamma$ for different system sizes $N$. 
For the non-Hermitian case we define the gap using the distance in the complex plane, see Eqs.~(\ref{d1},\ref{d2}).
Similarly, for the Hermitian case we define the gap as
\begin{equation}
  \Delta_{\rm H} =\max_i \left\{ \min_{j \ne i}\left[ \text{dist} \left(E_i, E_j\right) \right] \right\} \, ,
  \label{gaph2}
\end{equation}
where $\text{dist} \left(E_i, E_j\right)=|E_i-E_j|$ is the distance in the real axis (consistently with the non-Hermitian definition~\eqref{d2}). %, and we plot the analytical estimate~\eqref{gaph} for  comparison. 
With this definition, the presence of a finite and $N$ independent $\Delta_{\rm H,NH}$ in some region of $\gamma$ signals  the existence of an energy gap in the spectrum. In contrast, we have no energy gap in the spectrum in the region of parameters where $\Delta_{\rm H,NH}$ goes to zero as $N$ increases.

The continuous blue curve in Fig.~\ref{Fig:Gap} indicates the analytical estimate of the gap $\Delta_{\rm H,NH}$ for both cases: Eq.~\eqref{gaph} for the Hermitian case and Eq.~\eqref{gapnh} for the non-Hermitian one. The critical couplings $\gamma_{\rm cr}^{\rm H,NH}$ can be identified graphically as the values of $\gamma$ above which the numerical data for $\Delta_{\rm H,NH}$ (symbols) coincide with the analytical estimates (continuous blue curves). In the figure our predictions of the critical couplings given in Eq.~\eqref{gcrnhinf} (vertical dashed line in panel (a)) and Eq.~\eqref{gcrh} (arrows in panel (b)) are also shown.

Our analytical estimate for the gap works well above the critical  $\gamma$  for both Hermitian and non-Hermitian coupling, for all the
values of $N$ shown.
%The estimate of the critical point $\gamma_{\rm cr}^{H,NH}$ works well, too. 
  Interestingly we find that the critical $\gamma$ is independent of $N$ for large $N$ in the non-Hermitian case, as we predicted in Eq.~\eqref{gcrnhinf}, while it decreases with $N$ in the Hermitian case, according to our prediction~\eqref{gcrh}. Moreover from Fig.~\ref{Fig:Gap} one can see that the transition to a gapped phase in the non-Hermitian case is much sharper than the transition in the Hermitian case. 
  Note that for $\gamma<\gamma_{\rm cr}^{\rm H,NH}$ our estimate predicts that $\Delta_{\rm H,NH} \to 0$ for $\gamma\to 0$, while the numerical simulations show that $\Delta_{\rm H,NH} \to \delta=W/N$. This is clearly a  finite size effect  and it is not relevant since $\delta$ goes to zero when $N \to \infty$.
  
\section{Imaginary Energy gap and robustness to perturbations}
\label{secRob}

We have shown the emergence of both Hermitian and non-Hermitian gaps in the spectrum of a Picket-Fence model. While, in the Hermitian case, it is well known that a gap between the ground state and the excited states makes the first more robust to perturbations, it is not trivial that an imaginary gap has the same effect on the gapped state. Thus, here we will apply  perturbation theory to non-Hermitian systems and we  will show how the distance in the complex plane is related to the robustness to perturbations. Then we will show numerically how the non-Hermitian gap makes the system robust to static disorder.

\subsection{Non-Hermitian perturbative approach}

Let us consider a perturbation $D$ to the non-Hermitian Hamiltonian $H$, so that the total Hamiltonian of the system can be written as:
%given in Eq.~(\ref{H}), considering the non-Hermitian case, so that
\begin{equation}
\label{dis0}
H_D = H + D \, ,
\end{equation}
where $H$ is a generic non-Hermitian symmetric Hamiltonian.

Since $H$ is non-Hermitian, its eigenfunctions are not orthogonal. First of all, let us define a ``non-Hermitian bra'', being the transpose of a ket
\begin{equation}
\langle \bra{\psi} := (\ket{\psi})^t \, .
\end{equation}
Since the Hamiltonian is symmetric, the left eigenfunctions $\langle \bra{\psi_i}$ are the ``bra'' of the right eigenfunctions $\ket{\psi_i}$, that is
\begin{equation}
H\ket{\psi_i} ={\cal E}_i \ket{\psi_i} \quad \text{and} \quad \langle \bra{\psi_i} H = {\cal E}_i \langle \bra{\psi_i} .
\end{equation}
From here, the biorthogonality condition arises as
\begin{equation}
\langle \braket{\psi_i | \psi_j} = \delta_{ij} \ . 
\end{equation}

When the perturbation $D$ is sufficiently small, a perturbative correction of the complex eigenvalues up to second order can be derived \cite{Giulio}, and it has the expression
\begin{equation}\label{EigPert}
\overline{\cal E}_n \approx {\cal E}_n + \langle \braket{\psi_n | D | \psi_n}+\sum_{m \not= n}
 \frac{ \langle \braket{\psi_n | D | \psi_m}^2 }{{\cal E}_n - {\cal E}_m } \ .
\end{equation}
From Eq.~\eqref{EigPert} it is clear that the strength of the perturbation is determined by the ratio of two complex numbers $z_1 =\langle\braket{\psi_n | D | \psi_m}^2$, $z_2 ={\cal E}_n - {\cal E}_m$.
% which is small if the difference between the eigenenergies in the complex plane is much larger than the modulus of the number $z_1$.
This proves that a state separated by a gap in the complex plane from the rest of the spectrum is robust to perturbations as long as the gap is large compared to the modulus of the perturbations.

As a simple example of the above general calculations, 
let us consider a system made of two resonant sites, separated by a pure imaginary gap $i\gamma$, and perturbed with a coupling $D$. The corresponding non-Hermitian Hamiltonian is 
\[ H+D =
\left(
\begin{array}{cc}
  E_0 & 0  \\
 0 & E_0 -i\gamma
\end{array}
\right) +
\left(
\begin{array}{cc}
 0 & d  \\
 d & 0
\end{array}
\right)
\]
and the eigenenergies $\overline{\cal E}_\pm$ of $H+D$ can be analytically obtained as 
\begin{equation}\label{Eig0}
\overline{\cal E}_\pm=E_0-\frac{i\gamma}{2} \pm  \frac{i\gamma}{2} \sqrt{1 - \frac{4 d^2}{\gamma^2}} \, .
\end{equation}
Now, let us consider the case when $2d \ll \gamma$, i.e. the complex gap $\gamma$ is much larger than the coupling $d$ between the sites. Under this assumption we can expand the eigenenergies~\eqref{Eig0} to obtain
\begin{align} 
\overline{\cal E}_+ &\approx E_0 - \frac{i d^2}{\gamma} \label{EigExp1} \, , \\
\overline{\cal E}_- &\approx E_0 - i\gamma + \frac{i d^2}{\gamma} \label{EigExp2} \, .
\end{align}

The same result can be obtained by applying the perturbative expansion~\eqref{EigPert} and it shows that two unperturbed complex eigenenergies having the same real part but being distant in the imaginary axis can be robust to a perturbation, as long as the distance in the complex plane is much larger than the perturbation.

\subsection{Robustness of superradiance to static diagonal disorder}

\begin{figure}
    \centering
    \includegraphics[width=\columnwidth]{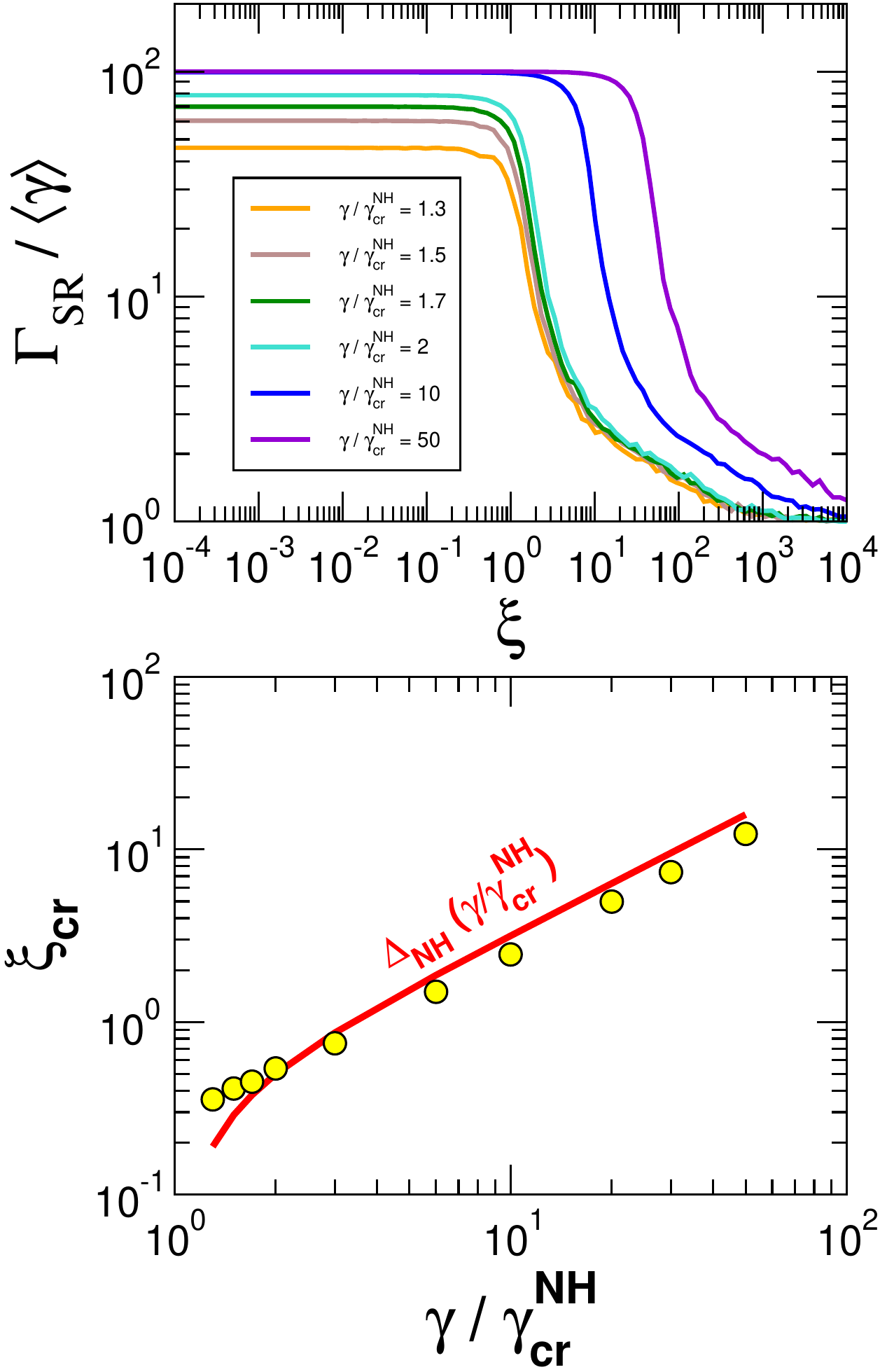}
    \caption{Upper panel: width of the superradiant state divided by the average width $\langle \gamma \rangle = \gamma / N$ {\it vs.} the disorder strength $\xi$ as introduced in the Hamiltonian~\eqref{dis0}-\eqref{dis1}-\eqref{dis2}. Lower panel: the critical value $\xi_{\rm cr}$ (see text) is plotted {\it vs.} $\gamma/\gamma_{\rm cr}^{\rm NH}$ as circles. The red line is the gap, as defined in Eq.~\eqref{gapnh}. Parameters in both panels are $N=100$, $W=1$ and an average over 100 realizations of static disorder is performed.}
    \label{robust}
\end{figure}

In order to check that the previous results are valid in the model considered here beyond the perturbative regime, let us  add to the 
the SES Hamiltonian 
\begin{equation}
    \label{dis1}
    H = \sum_k E_k \ket{k} \bra{k} - i\frac{\gamma}{2N} \sum_{k,k'} \ket{k}
  \bra{k'},
\end{equation}
the static disorder
\begin{equation}
\label{dis2}
    D = \sum_k \epsilon_k \ket{k} \bra{k} \, ,
\end{equation}
where $\epsilon_k$ are random numbers uniformly distributed such that $\epsilon_k \in [-\xi/2, \xi/2]$. Here the parameter $\xi$ is proportional to the standard deviation of the energy fluctuations introduced by $D$ and it represents the disorder strength.
In particular, our aim is     to 
  study the robustness of the superradiant state of the non-Hermitian case   to such static disorder.  
 In Fig.~\ref{robust} (upper panel) the width of the superradiant state $\Gamma_{\rm SR}$ divided by the average width $\langle \gamma \rangle = \gamma/N$ is shown {\it vs.} the disorder strength $\xi$ for different values of $\gamma$ larger than the critical $\gamma_{\rm cr}^{\rm NH}$. 
As one can see, the width of the superradiant state is larger than $\langle \gamma \rangle$ for small disorder $\xi$. Then, beyond some critical value of $\xi$, the width start to decrease with $\xi$, ultimately reaching $\Gamma_{\rm SR} = \langle \gamma \rangle = \gamma/N$ for $\xi \to \infty$. 
In order to quantify phenomenologically such critical disorder strength, let us define
  a  critical value $\xi_{\rm cr}$ as the value of $\xi$ beyond which the width of the superradiant state is less than 95\% of its value without disorder. In this sense, $\xi_{\rm cr}$ is proportional to the disorder strength needed to destroy superradiance. In the lower panel of Fig.~\ref{robust} $\xi_{\rm cr}$ is plotted {\it vs.} the ratio $\gamma/\gamma_{\rm cr}^{\rm NH}$. In the same panel, the gap~\eqref{gapnh} is plotted as a comparison. As one can see, apart from small deviations where $\gamma \simeq \gamma_{\rm cr}^{\rm NH}$, the critical disorder $\xi_{\rm cr}$ increases with $\gamma$ and it is approximately proportional to the non-Hermitian gap. This shows that the non-Hermitian gap makes superradiance robust to static disorder.

\section{Relation between the gapped regime and the interaction range}
\label{secAlpha}
 
In this section we want to extend our analysis to different ranges of interaction. In particular, we want to know if the emergence of the gap is a general outcome of long-range interactions in a PF model.
 Differently from the rest of the paper, here we focus only on the Hermitian case. 
Note that the Hermitian model is relevant in many realistic situations, such as ion traps~\cite{iontraps,robincs}, where one-dimensional systems with tunable interaction range can be emulated.

We model an interaction of range $\alpha$ by the Hamiltonian term
\begin{equation}
V = -\frac{\gamma}{2v_{N,\alpha}}\sum_{\substack{k,k'\\k \neq k'}} \frac{\ket{k} \bra{k'}}{|k-k'|^\alpha} \, ,
\label{Valpha}
\end{equation}
where $v_{N,\alpha}$ is a normalization constant and, since we are dealing with a one-dimensional system, we speak about ``long-range interaction'' for $0\leq\alpha<1$ and about ``short-range interaction'' for $\alpha>1$. Note that in ion trap experiments~\cite{iontraps} the exponent $\alpha$ can be tuned from 0 to 3. The case $\alpha=1$ is critical because $\alpha$ equals the dimension of the system and thus we will analyze it separately. The normalization constant $v_{N,\alpha}$ has been added 
in order to have an extensive Hamiltonian energy and to fix 
the spectrum of $V$ as large as  $\gamma/2$. For the case $\alpha=0$, for example, we have $v_{N,0}=N$, which is exactly the Hermitian case studied in the previous Sections. For $\alpha\neq 0$, $v_{N,\alpha}$ is determined by numerically diagonalizing $V$, and it has the following scaling with the system size (see Appendix~\ref{appAlpha} for details):
\begin{equation}
    v_{N,\alpha} \sim
    \begin{cases}
        N^{1-\alpha} & \text{for} \, \alpha < 1 \\
        \ln N & \text{for} \, \alpha = 1 \\
        \text{const.} & \text{for} \, \alpha > 1
    \end{cases}
    \, .
    \label{vNalpha}
\end{equation}

In order to understand how the presence of a gap is  connected to the range of the interaction, here we study numerically the presence of the gap $\Delta_{\rm H}$ defined in Eq.~\eqref{gaph2}. Let us remind that the presence of a finite and $N$-independent $\Delta_{\rm H}$ in some region of $\gamma$ signals  the existence of an energy gap in the spectrum.

\begin{figure}[!ht]
  \centering
  \includegraphics[width=\columnwidth]{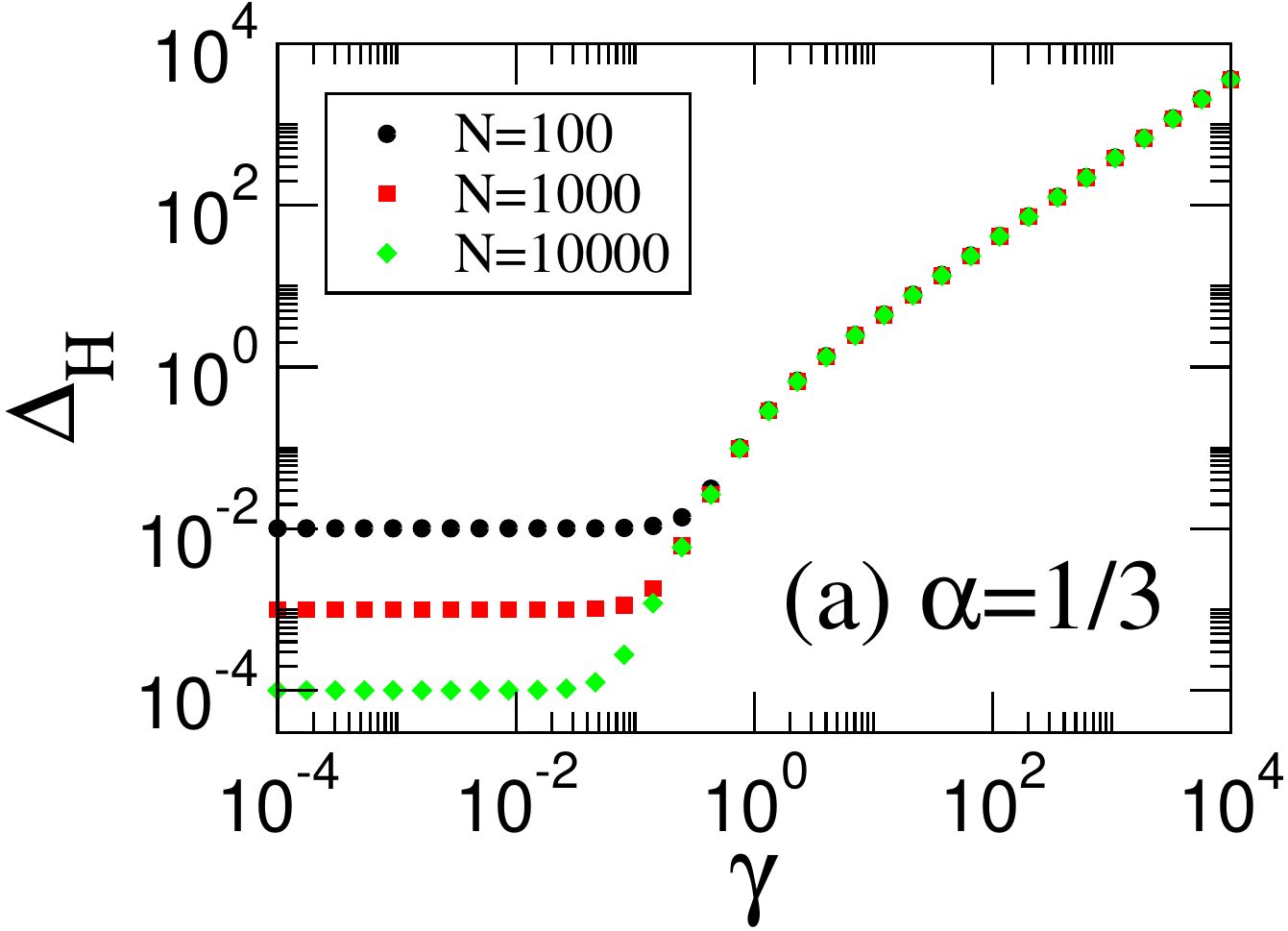}
   \includegraphics[width=\columnwidth]{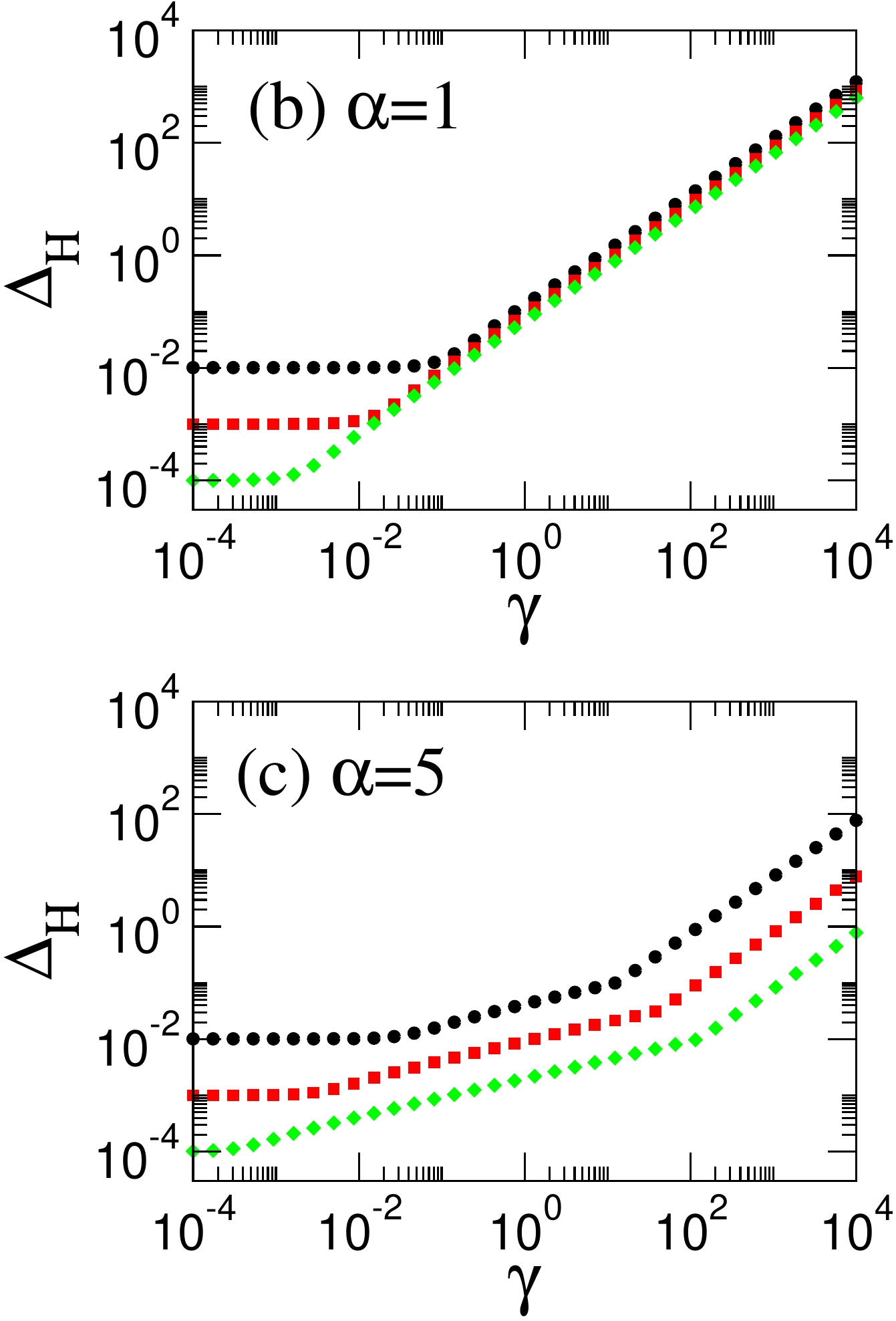}
  \caption{Gap $\Delta_{\rm H}$, as defined in~\eqref{gaph2}, {\it vs.}~the coupling parameter $\gamma$ with the interaction~\eqref{Valpha}. Here $W=1$ and $N=\{100,1000,10000\}$.
}
  \label{Fig:alpha}
\end{figure}

In Fig.~\ref{Fig:alpha} we plot $\Delta_{\rm H}$, defined as in Eq.~\eqref{gaph2}, as a function of $\gamma$ for $\alpha=\{1/3,1,5\}$. The case $\alpha=1/3$ shown in panel (a) corresponds to a long-range interaction and one can see that, similarly to the case $\alpha=0$ (see Fig.~\ref{Fig:Gap}(b)), the gap is independent of the system size $N$ for large $\gamma$. On the other hand, for short-range $\alpha=5$ (panel (c)) $\Delta_{\rm H}$ decreases with the system size for any value of $\gamma$ and thus there is no gap for these two cases in the limit $N \to \infty$.
For the critical range $\alpha=1$ (panel (b)) the results are less clear and more analysis is needed to establish the non-existence of a gapped regime (as the data shown in the Figure seem to indicate).

\begin{figure}[ht]
    \centering
    \includegraphics[width=0.5\textwidth]{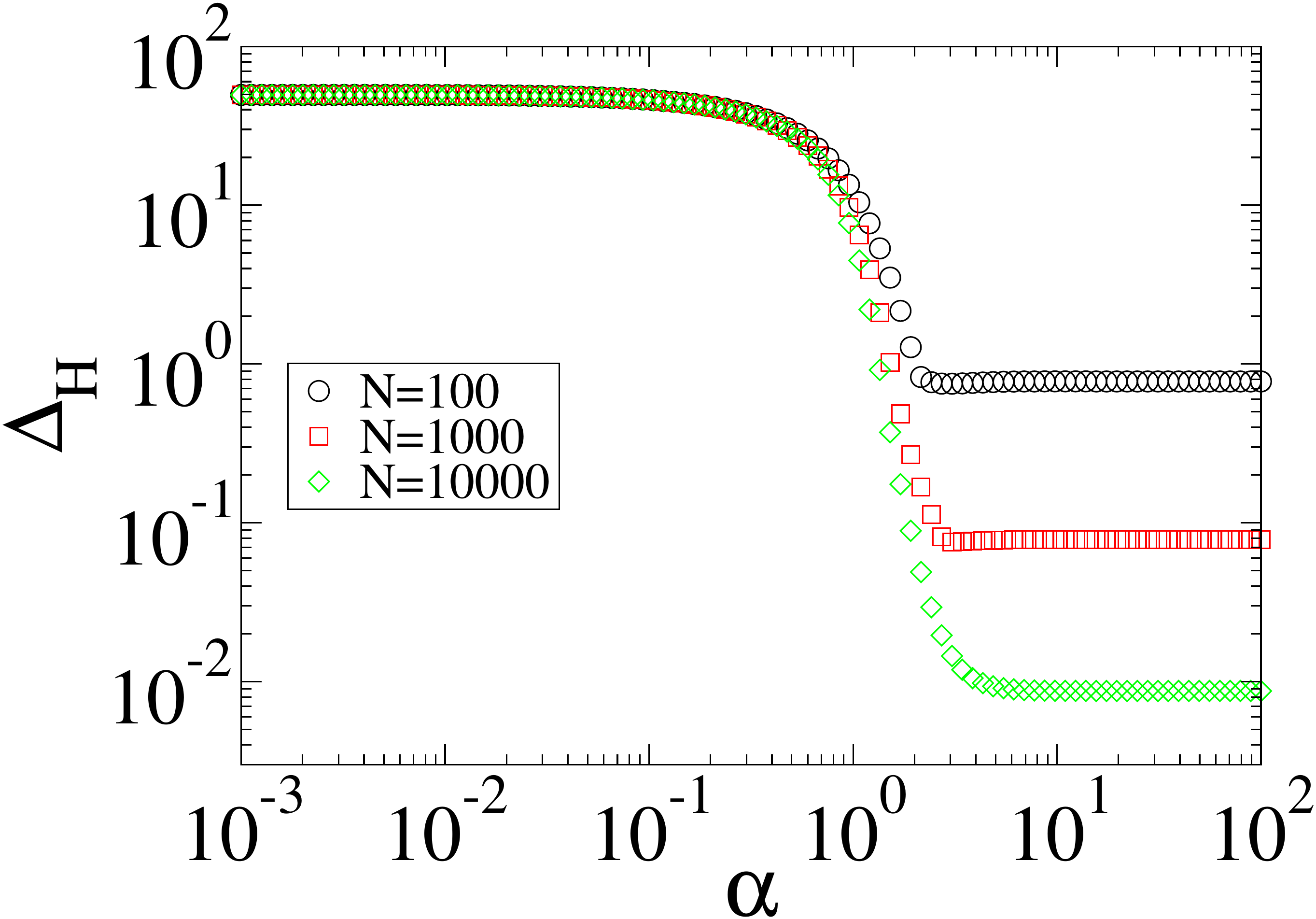}
    \caption{Gap $\Delta_{\rm H}$ {\it vs.}~the range of interaction $\alpha$ for N=\{100,1000,10000\}, where the symbols are given by Eq.~\eqref{gaph2}. Here, $W=1$ and $\gamma=100$.}
    \label{fig:Gapalpha}
\end{figure}

From Fig.~\ref{Fig:alpha} one can see that $\gamma=100$ represents a ``strong-coupling'' regime for the three values of $\alpha$ shown. Then, in Fig.~\ref{fig:Gapalpha} we plot $\Delta_{\rm H}$ (symbols) {\it vs.} $\alpha$ for $\gamma=100$ setting the same parameters and the same values of $N$ as in Fig.~\ref{Fig:alpha}. In Fig.~\ref{fig:Gapalpha} one can see two different regimes: (i) for $\alpha < 1$ (long range)) $\Delta_{\rm H}$ is independent of $N$, representing a gap in the $N \to \infty$ limit; (ii) for $\alpha \gtrsim 1$ (short range) $\Delta_{\rm H}$ decreases with $N$, meaning that there is no gap in the thermodynamic limit. Note that we checked that in the long range regime ($\alpha < 1$)  the gap arises between the ground state and the first excited state, i.e. $\Delta_{\rm H}=E_2-E_1$.

As a final remark we note that in this Section
 we have analyzed the role of the interaction range just in the Hermitian case, because 
adding a variable range of the interaction in the non-Hermitian case is more difficult. 
 The generalization to different ranges that we used for the Hermitian case, in fact, cannot be performed as it is for the non-Hermitian case without loss of consistency. Indeed, an imaginary interaction $V_{k,k'}=-i\gamma|k-k'|^{-\alpha}$  would lead to both positive and negative decay widths, while the decay widths of a non-Hermitian Hamiltonian are required to be all positive. Nevertheless let us note that in realistic molecular system the non-Hermitian interaction can have a complicated power law decay with the distance~\cite{photosynthesis}.

\section{Conclusions and Perspectives}

We have compared two paradigmatic models for Superconductivity and
Superradiance, focusing on the emergence of an energy gap
in the real and complex plane, respectively. We show that a gap arises
also in the Superradiance model in the complex plane, and we give an
analytical estimate of that gap which agrees very well with our
numerical simulations. We show that the usual Superradiance transition
can be interpreted as a transition to a gapped regime. Moreover in the
large coupling limit the Superradiance  and the Superconductivity gaps are 
mathematically the
same, while they differ at the criticality. Indeed, while
the critical value for the emergence of Superconductivity depends on
the system size, the critical value for the emergence of Superradiance is
independent of the system size. 
Finally we have also shown that a gap in the complex plane can induce robustness to perturbations in the system, similarly to a gap in the real axis. 
This result allows to interpret the robustness of Superradiance to disorder reported in several previous publications in literature as a consequence of the presence of an imaginary energy gap. 
In perspective, the relevance of these energy gaps to transport and
other system properties will be analyzed. 
From a mathematical point of view, we have shown that 
the emergence of such gapped states can be connected with 
the long range nature of the interaction. Indeed both the discrete BCS model and the SES model share a distance  independent coupling (all-to-all coupling). The connection of a gapped state with the long range of the interaction has been also pointed out in Ref.~\cite{robincs} by some of the Authors of this paper.

Even if here we have analyzed very simple models amenable of analytic treatment, our results can be relevant for a variety of realistic systems, such as molecular chains~\cite{mchains}, ion traps~\cite{iontraps} and photosynthetic systems~\cite{photosynthesis}. For instance the range of interaction can be controlled in ion trap experiments~\cite{iontraps}  where the Hermitian Hamiltonian discussed here  can be experimentally realized. Moreover a linear molecular chain interacting with an electromagnetic field can be modelled by  non-Hermitian Hamiltonians~\cite{photosynthesis} very similar to the ones considered here. 
In perspectives we plan to analyze realistic models for quantum transport in presence of non-Hermitian and Hermitian interactions and to study the relevance of the gapped regime to the efficiency of transport. We expect that the existence of gapped extended states can act as a support for efficient energy transport.  

%We expect such gapped regime
%to deeply influence the transport properties. 

% \section{Acknowledgment}
\begin{acknowledgement}

GLC acknowledges Emil Yuzbashyan for pointing out the similarity between  the discrete BCS model and the SES model and the Center for Theoretical Physics of Complex Systems (PCS-IBS) in Daejeon, South Korea, where this work began, for the hospitality.
This work was partially supported by 
VIEP-BUAP (Grant Nos.~MEBJ-EXC18-I and LC-EXC18-G), 
Fondo Institucional PIFCA (Grant No.~BUAP-CA-169), and 
CONACyT (Grant No.~CB-2013/220624).
GLC acknowledges  the support of PRODEP (511-6/17-8017).
N.C.C. acknowledges the support of a scholarship from CONACyT and the hospitality of 
the Department of Mathematics and Physics of the Catholic University of the Sacred Heart, Italy, where part of this work has been done.

\end{acknowledgement}

\appendix

\section{Non-Hermitian coupling: widths of the subradiant states}
\label{appsub}

In the main text, discussing the case of a non-Hermitian coupling, we report the analytic expression of the widths of all the eigenstates for $\gamma < \gamma_{\rm SR}$~\eqref{width3} and of the widths of the subradiant states for $\gamma > \gamma_{\rm SR}$~\eqref{width4}. Here we derive those expressions, as well as the critical coupling $\gamma_{\rm SR}$, following~\cite{rotter}.

Let us consider the case of odd $N$, so that we can write $N=2M+1$ with $M$ an integer. Note that the limit $N \to \infty$ corresponds to $M \to \infty$ and in that limit there is no distinction between even or odd values of $N$. The Hamiltonian~\eqref{H}, with $E_k$ given by~\eqref{pf} and $V_0=i\gamma/(2N)$, can be mapped to
\begin{equation}
  H = \delta \overline{H} = \delta \left( \sum_{k=-M}^M k \ket{k}\bra{k} - i \alpha \sum_{k=-M}^M \sum_{k'=-M}^M \ket{k}\bra{k'} \right) \, ,
\end{equation}
where the center of the unperturbed spectrum is assumed to be at $E_0=0$, without loss of generality, and the coupling parameter is
\begin{equation}
  \label{alpha}
  \alpha = \frac{\gamma}{2N\delta} = \frac{\gamma}{2W} \, .
\end{equation}

We now proceed to compute the eigenvalues $\overline{\lambda}$ of $\overline{H}$, which are related to the eigenvalues $\lambda$ of $H$ by $\lambda = \delta \overline{\lambda}$. Thus, let us consider the matrix $\bra{k} \overline{H} \ket{k'}$. By construction, all column and row vectors, respectively, of the non-Hermitian part of that matrix are linearly dependent. Summing $i\alpha$ times the central row ($k = 0$) to all the other rows ($k \ne 0$), one gets the following expression for the characteristic polynomial:
\begin{equation}\label{eq1:Rotter}
  P_M(\overline{\lambda}) = \displaystyle\prod_{k=-M}^M (k-\overline{\lambda})-i\alpha \sum _{k=-M}^M \prod _{\substack{j=-M \\ j\neq k}}^M (j-\overline{\lambda} )=0 \, .
\end{equation}
According to Eq.~\eqref{eq1:Rotter}, $P_M(\overline{\lambda})$ is the sum of two polynomials,
\begin{equation}\label{eq2:Rotter}
  P_M(\overline{\lambda})=Q_M(\overline{\lambda})-i\alpha R_M(\overline{\lambda})
\end{equation}
which are related in a simple manner,
\begin{equation*}
  R_M=-\frac{d}{d\overline{\lambda}}Q_M \, .
\end{equation*}
Taking the limit $M \to \infty$ and using the infinite product expansion of the sine function we have
\begin{equation}
  P_\infty(\overline{\lambda})=\sin[\pi \overline{\lambda} ]+i \alpha  \pi  \cos[\pi \overline{\lambda} ]=0
\end{equation}
with $\overline{\lambda}=\overline{E}-\frac{i}{2}\overline{\Gamma}$. Then, we can substitute this expression of $\overline{\lambda}$ to get
\begin{equation}
  P_\infty(\overline{\lambda} )=\sin\left[\pi \left(\overline{E} -\frac{i}{2}\overline{\Gamma} \right)\right]+i \alpha  \pi  \cos\left[\pi \left(\overline{E} -\frac{i}{2}\overline{\Gamma} \right)\right]=0 \, .
\end{equation}
Separating real and imaginary parts one has
\begin{equation}
  \begin{cases}
    \sin[\pi \overline{E} ] \left[e^{\pi  \overline{\Gamma} }(1-\alpha  \pi )+(1+\alpha  \pi )\right]=0\\
    \cos[\pi \overline{E} ]\left[e^{\pi  \overline{\Gamma} }(1-\alpha  \pi )-(1+\alpha  \pi )\right]=0
  \end{cases} .
\end{equation}
So there are two solutions:\\
i) $\sin[\pi \overline{E} ]=0$, $\overline{E}=n\in \mathbb{Z}$
\begin{equation}
  e^{\pi  \overline{\Gamma} }=\frac{1+\alpha  \pi }{1-\alpha  \pi } \, ,
\end{equation}
from which
\begin{equation}\label{Rotter1}
  \overline{\Gamma}=\frac{1}{\pi}\ln \left(\frac{1+\alpha  \pi }{1-\alpha  \pi }\right)
\end{equation}
under the conditions $e^{\pi  \overline{\Gamma} }>0$, $\alpha<\frac{1}{\pi}$. This result represents the widths of all the eigenstates below the Superradiance transition.\\
ii) $\cos[\pi \overline{E} ]=0$, $\overline{E}=n+\frac{1}{2}, n \in \mathbb{Z}$
\begin{equation}
  e^{\pi  \overline{\Gamma} }=\frac{\alpha  \pi +1}{\alpha  \pi -1} \, ,
\end{equation}
which gives
\begin{equation}\label{Rotter2}
  \overline{\Gamma}_{\rm sub}=\frac{1}{\pi}\ln\left(\frac{\alpha  \pi +1}{\alpha  \pi -1}\right)
\end{equation}
under the conditions $e^{\pi  \overline{\Gamma} }>0$, $\alpha>\frac{1}{\pi}$. This result represents instead the widths of the subradiant states above the Superradiance transition. From these results a critical coupling parameter $\alpha_\mathrm{SR}=1/\pi$ emerges, which marks the Superradiance transition.

Now, let us map our expression for $\overline{\lambda}$ to $\lambda=E-\frac{i}{2}\Gamma$. Multiplying by $\delta$ we have
\begin{subequations}
  \begin{align}
    E_n &= n \delta \qquad (n \in \mathbb{Z}) \\ 
    \Gamma&=\frac{\delta}{\pi}\ln\left(\frac{1+\alpha/\alpha_{\rm SR}}{1-\alpha/\alpha_{\rm SR} }\right) \qquad &\text{for } \alpha < \alpha_{\rm SR} \label{width1}
  \end{align}
\end{subequations}
and
\begin{subequations}
  \begin{align}
    E_n &= \left( n + \frac{1}{2} \right) \delta \qquad (n \in \mathbb{Z}) \\
    \Gamma_{\rm sub}&=\frac{\delta}{\pi}\ln\left(\frac{\alpha/\alpha_{\rm SR} +1}{\alpha/\alpha_{\rm SR} -1}\right) \qquad &\text{for } \alpha > \alpha_{\rm SR} \, , \label{width2}
  \end{align}
\end{subequations}
where we can use~\eqref{alpha} to express the ratio between $\alpha$ and  $\alpha_\mathrm{SR}$ as
\begin{equation}
  \frac{\alpha}{\alpha_\mathrm{SR}} = \frac{\gamma \pi}{2W} \, .
\end{equation}
Thus, equations (\ref{width1}) and (\ref{width2}) can be rewritten in terms of the parameters of $H$ as
\begin{equation}\label{width3a}
  \Gamma=\frac{W}{N\pi}\ln\left(\frac{1+\gamma/\gamma_{\rm SR}}{1-\gamma/\gamma_{\rm SR}}\right) \qquad \text{for } \gamma < \gamma_{\rm SR}
\end{equation}
and
\begin{equation}\label{width4a}
  \Gamma_{\rm sub}=\frac{W}{N\pi}\ln\left(\frac{\gamma/\gamma_{\rm SR} +1}{\gamma/\gamma_{\rm SR} -1}\right) \qquad \text{for } \gamma > \gamma_{\rm SR} \, ,
\end{equation}
by defining the critical coupling
\begin{equation}
  \gamma_\mathrm{SR}=\frac{2W}{\pi} \, .
\end{equation}

\section{Long and short-range interaction}
\label{appAlpha}

\begin{figure*}[t]
  \centering
  \includegraphics[width=0.9\textwidth]{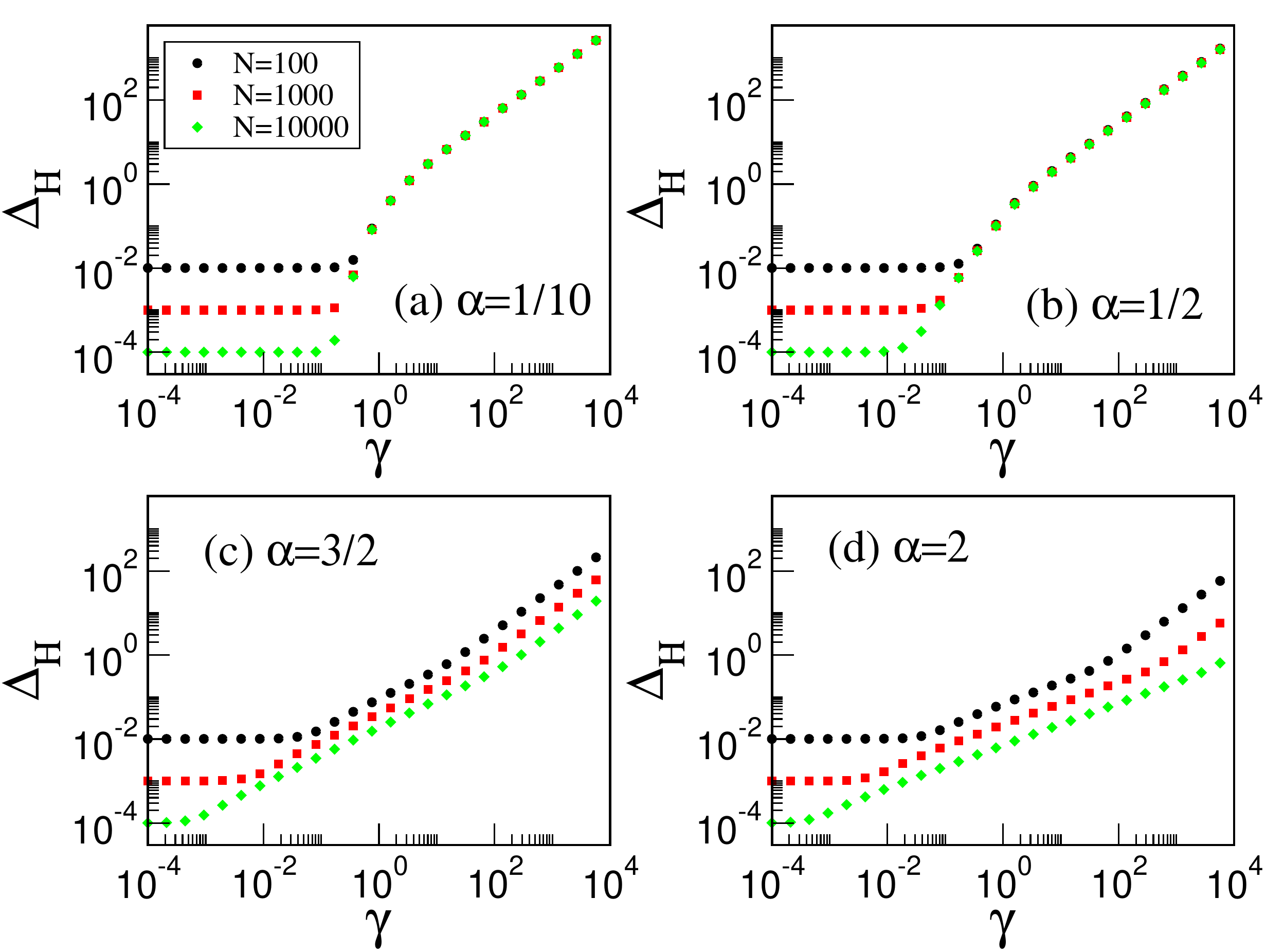}
  \caption{a) Gap $\Delta_{\rm H}$, as given by Eq.~\eqref{gaph2}, {\it vs.}~the coupling $\gamma$ with the interaction~\eqref{Valpha} for different values of the interaction range $\alpha$.
  %=\{1/10,1/2,3/2,2\}$. 
  Here we set $W=1$.
  %and $N=\{100,1000,10000\}$. Symbols are  while full lines represent $E_2 - E_1$.
}  \label{Fig:alpha1}
\end{figure*}

In the text, we reported how the gap $\Delta_{\rm H}$ changes with the range of the interaction for $\alpha=\{1/3,1,5\}$. Here, in Fig.~\ref{Fig:alpha1}, we show the dependence of $\Delta_{\rm H}$ on $\gamma$ for some additional values of the range of interaction, namely for $\alpha=\{1/10,1/2,3/2,2\}$. 
%We also compare the definition~\eqref{gaph2} (symbols) with $E_2 - E_1$ (curves), showing that the two are equivalent for $\alpha \leq 2$, in the range of $\gamma$ that we plotted.
We would like to point out also that the definition~\eqref{gaph2} is equal to $E_2 - E_1$ in the range of $\gamma$ that we plotted in this figure and in the main text (Fig.~\ref{Fig:alpha}).

%\begin{figure}[ht]
 % \centering
 % \includegraphics[width=1\columnwidth]{fig/prPFalpha-0}
 % \caption{a) Gap $\Delta_{\rm H}$, {\it vs.}~the coupling parameter $\gamma$. Here, $\alpha=0$, $W=100$ and $N=\{40,160,1000,4000,10000\}$.}
%  \label{Fig:alpha0}
%\end{figure}

\begin{figure}[ht]
  \centering
  \includegraphics[width=0.5\textwidth]{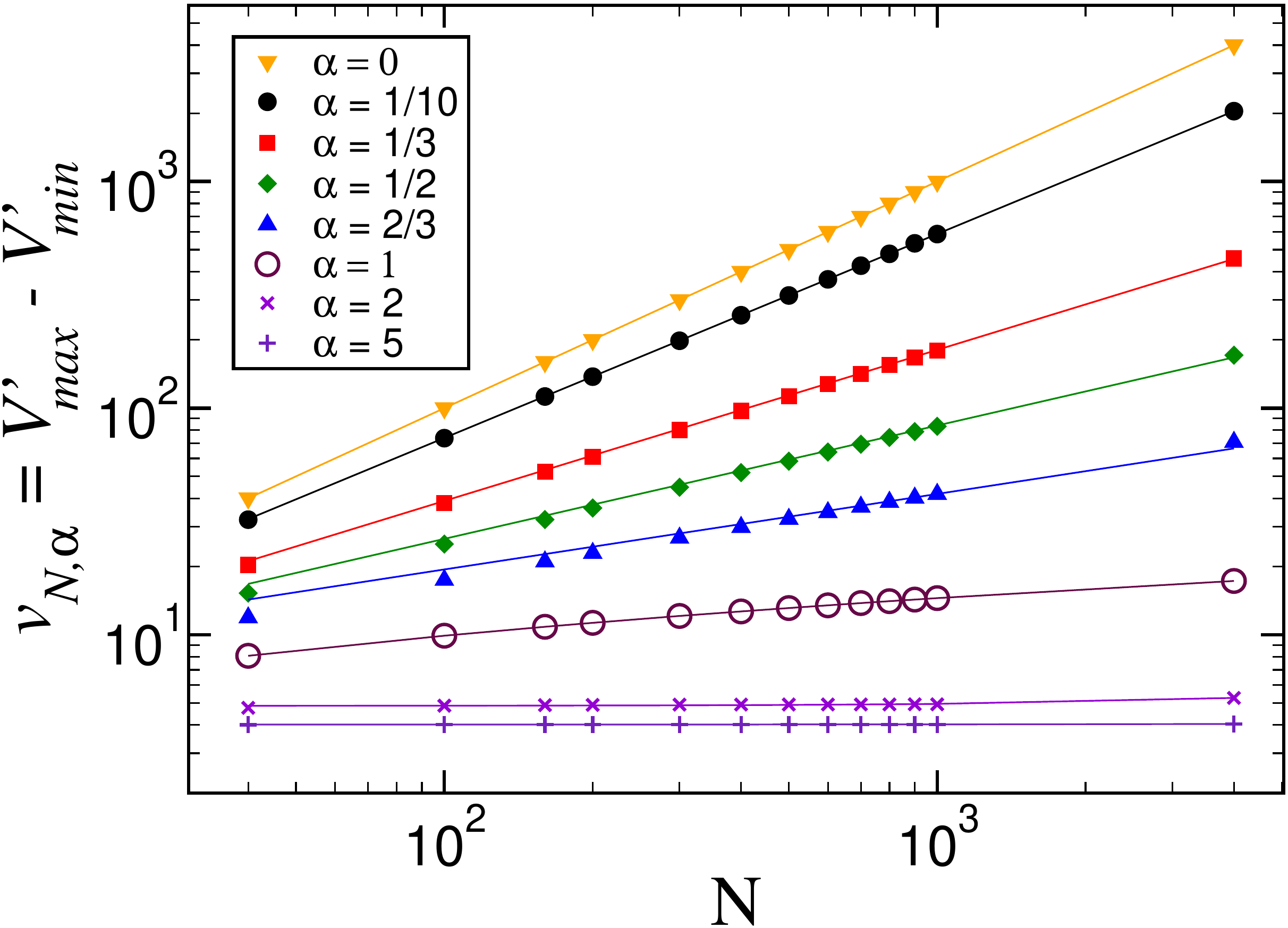}
  \caption{The energy range of the interaction $v_{N,\alpha}$ {\it vs.} the system size $N$ for different $\alpha=\{0,1/10,1/3,1/2,2/3,1,2,5\}$. Symbols are obtained by diagonalizing $V$~\eqref{Valpha} without the prefactor $\gamma/(2v_{N,\alpha})$, while lines are the best fits with the functions~\eqref{vNalpha}.}
  \label{Fig:V}
\end{figure}

As we reported in the Sec.~\ref{secAlpha}, in order to obtain the gap $\Delta_{\rm H}$ for different range of interaction $\alpha$, the interaction~\eqref{Valpha} is normalized by the constant $v_{N,\alpha}$ defined as the difference between the maximum eigenenergy and minimal eigenenergy of the matrix $V$ given in Eq.~\eqref{Valpha} without the prefactor ($\gamma/(2v_{N,\alpha})$), i.e. 
$v_{N,\alpha}=V_{max}-V_{min}$. In Fig.~\ref{Fig:V} we plot $v_{N,\alpha}$ vs. $N$ for different values of $\alpha$ and we show that the exact results obtained from the diagonalization of $V$ (symbols) fit well the scaling~\eqref{vNalpha} for all the values of $\alpha$ shown here.

\begin{thebibliography}{9}
\bibitem{note}
In the BCS theory~\cite{schrieffer} and in Ref.s~\cite{bcs,robincs} the following expression for the gap is reported: $\Delta_{\rm BCS} = W / \sinh (2W/\gamma)$, which gives different results from Eq.~\eqref{gaph} in some parameter range. In particular, when $W \gg \gamma/2$, Eq.~\eqref{gaph} and $\Delta_{\rm BCS}$ differ by a factor of 2, namely $\Delta_{\rm H} \approx W \, e^{-2W/\gamma}$ and $\Delta_{\rm BCS} \approx 2W \, e^{-2W/\gamma}$. On the other hand, in the opposite limit $W \ll \gamma/2$, both~\eqref{gaph} and $\Delta_{\rm BCS}$ have the same approximated expression $\Delta_{\rm H} \approx \Delta_{\rm BCS} \approx \gamma / 2$. Nevertheless, both expressions predict that the gap closes in the limit $\gamma/W \to 0$.
\end{thebibliography}
\end{document}